\documentclass[%
 reprint,
 groupedaddress,
 showpacs, % can also use noshowpacs to turn this off
 noshowkeys,% can also use noshowkeys to turn this off
 amsmath,amssymb,
 aps,
 prl,
]{revtex4-1}

\usepackage{graphicx}% Include figure files
\usepackage{dcolumn}% Align table columns on decimal point
\usepackage{bm}% bold math
\usepackage[cdot,mediumqspace,amssymb]{SIunits} % type dimensionful numbers as e.g. \unit{5.1}{\micro\metre}
\usepackage[autolanguage]{numprint} % use to print numbers, e.g. \numprint{1.23456E-1}

\usepackage{ifthen}
\usepackage{ifpdf}
\ifpdf
    \graphicspath{{Figures/PDF/}}
\else
    \graphicspath{{Figures/EPS/}}
\fi

\usepackage{cleveref}
\usepackage{color}
\begin{document}

\title{Robust Detection of Dynamic Community Structure in Networks}

\author{Danielle S. Bassett$^{1,2,*}$, Mason A. Porter$^{3,4}$, Nicholas F.
Wymbs$^{5}$, Scott T. Grafton$^{5}$, Jean M. Carlson$^{1}$, Peter J.
Mucha$^{6,7}$} \affiliation{$^1$Department of Physics, University of
California, Santa Barbara, CA 93106, USA;\\ $^{2}$ Sage Center for the Study of
the Mind, University of California, Santa Barbara, CA 93106;\\ $^{3}$ Oxford
Centre for Industrial and Applied Mathematics, Mathematical Institute,
University of Oxford, Oxford OX1 3LB, UK;\\ $^{4}$CABDyN Complexity Centre,
University of Oxford, Oxford, OX1 1HP, UK;\\ $^{5}$Department of Psychology and
UCSB Brain Imaging Center, University of California, Santa Barbara, CA 93106,
USA;\\ $^{6}$Carolina Center for Interdisciplinary Applied Mathematics,
Department of Mathematics, University of North Carolina, Chapel Hill, NC
27599, USA;\\ $^{7}$Institute for Advanced Materials, Nanoscience \&
Technology, University of North Carolina, Chapel Hill, NC 27599, USA;\\ $^*$Corresponding author. Email address: dbassett@physics.ucsb.edu}

\date{\today}

%%%%%%%

\begin{abstract}

We describe techniques for the robust detection of community structure in
some classes of time-dependent networks. Specifically, we consider the use of
statistical null models for facilitating the principled identification of
structural modules in semi-decomposable systems. Null models play an
important role both in the optimization of quality functions such as
modularity and in the subsequent assessment of the statistical validity of
identified community structure. We examine the sensitivity of such methods to
model parameters and show how comparisons to null models can help identify
system scales.  By considering a large number of optimizations, we quantify
the variance of network diagnostics over optimizations (`optimization
variance') and over randomizations of network structure (`randomization
variance').  Because the modularity quality function typically has a large
number of nearly-degenerate local optima for networks constructed using real
data, we develop a method to construct representative partitions that uses a
null model to correct for statistical noise in sets of partitions. To
illustrate our results, we employ ensembles of time-dependent networks
extracted from both nonlinear oscillators and empirical neuroscience data.
\end{abstract}

% PACS, the Physics and Astronomy
                             % Classification Scheme.

%Use showkeys class option if keyword
                              %display desired
\maketitle

%%%%%%%%%%%%%

%%%%%%%%%%%%%%

{\bf Many social, physical, technological, and biological systems can be
modeled as networks composed of numerous interacting parts \cite{Newman2010}.
As an increasing amount of time-resolved data has become available, it has
become increasingly important to develop methods to quantify and characterize
dynamic properties of \emph{temporal networks} \cite{Holme2011}. Generalizing
the study of static networks, which are typically represented using graphs,
to temporal networks entails the consideration of nodes (representing
entities) and/or edges (representing ties between entities) that vary in
time.  As one considers data with more complicated structures, the
appropriate network analyses must become increasingly nuanced.  In the
present paper, we discuss methods for algorithmic detection of dense clusters
of nodes (i.e., communities) by optimizing quality functions on multilayer
network representations of temporal networks \cite{Mucha2010,Bassett2011b}.
We emphasize the development and analysis of different types of null-model
networks, whose appropriateness depends on the structure of the networks one
is studying as well as the construction of representative partitions that
take advantage of a multilayer network framework.  To illustrate our ideas,
we use ensembles of time-dependent networks from the human brain and human
behavior.}

\section{Introduction}

Myriad systems have components whose interactions (or the components
themselves) change as a function of time.  Many of these systems can be
investigated using the framework of \emph{temporal networks}, which consist
of sets of nodes and/or edges that vary in time \cite{Holme2011}.  The
formalism of temporal networks is convenient for studying data drawn from
areas such as person-to-person communication (e.g., via mobile phones
\cite{Onnela2007,Wu2010}), one-to-many information dissemination (such as
Twitter networks \cite{morenotwitter}), cell biology, distributed computing,
infrastructure networks, neural and brain networks, and ecological networks
\cite{Holme2011}. Important phenomena that can be studied in this framework
include network constraints on gang and criminal activity
\cite{Fararo1997,andrea2011}, political processes \cite{Mucha2010b,waugh09},
human brain function \cite{Bassett2011b,Doron2012}, human behavior
\cite{Wymbs2011}, and financial structures \cite{Fenn2009,Fenn2011}.

\begin{figure}[]
\includegraphics[width=.4\textwidth]{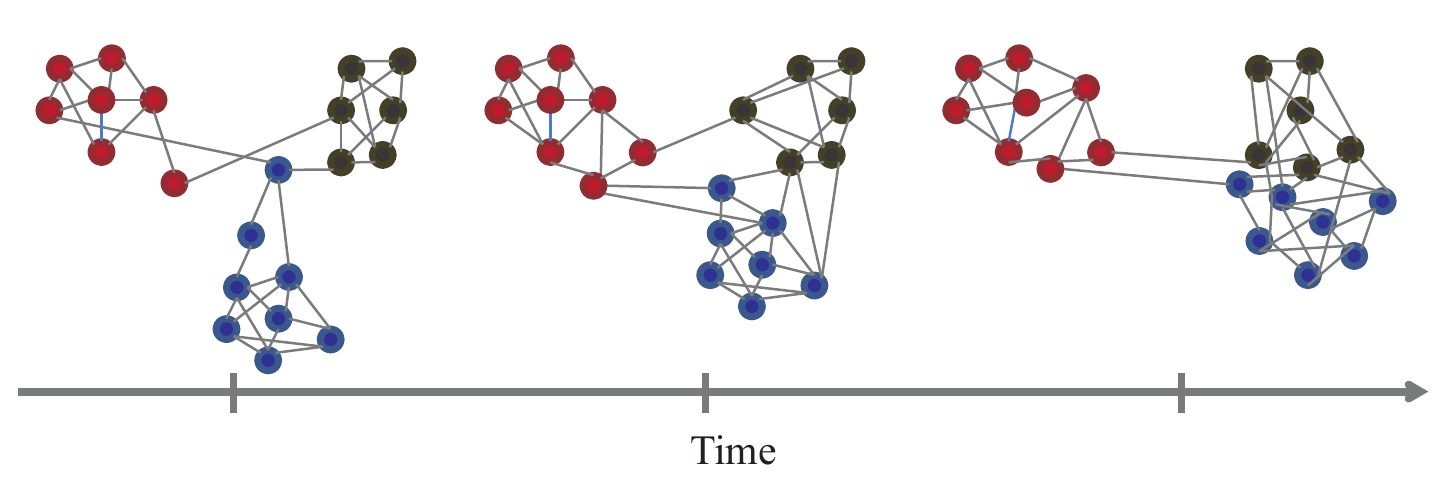}
\caption[]{(Color Online) An important property of many real-world networks
is community structure, in which there exist cohesive groups of nodes such that a network has stronger connections within such groups than it does between such groups. Community structure often changes in time, which can lead to the rearrangement of cohesive groups, the formation of new groups, and the breakup of existing groups. \label{Fig0}}
\end{figure}

Time-dependent complex systems can have densely connected components in the
form of cohesive groups of nodes known as `communities' (see
Fig.~\ref{Fig0}), which can be related to a system's functional modules
\cite{Porter2009,herbert}. A wide variety of clustering techniques have been
developed to identify communities, and they have yielded insights in the
study of the committee structure in the United States Congress
\cite{congshort}, functional groups in protein interaction networks
\cite{anna}, functional modules in brain networks \cite{Bassett2011b}, and
more. A particularly successful technique for identifying communities in
networks \cite{Porter2009,Fortunato2010} is optimization of a quality
function known as `modularity' \cite{NG2004}, which recently has been
generalized for detecting communities in time-dependent and multiplex
networks \cite{Mucha2010}.

Modularity optimization allows one to algorithmically partition a network's
nodes into communities such that the total connection strength within groups
of the partition is more than would be expected in some null model.  However,
modularity optimization always yields a network partition (into a set of
communities) as an output whether or not a given network truly contains
modular structure.  Therefore, application of subsequent diagnostics to a
network partition is potentially meaningless without some comparison to
benchmark or null-model networks.  That is, it is important to establish
whether the partition(s) obtained appear to represent meaningful community
structures within the network data or whether they might have reasonably
arisen at random. Moreover, robust assessment of network organization depends
fundamentally on the development of statistical techniques to compare
structures in a network derived from real data to those in appropriate models
(see, e.g., Ref.~\cite{Lancichinetti2010}).  Indeed, as the constraints in
null models and network benchmarks become more stringent, it can become
possible to make stronger claims when interpreting organizational structures
such as community structure.

\begin{figure}[]
\includegraphics[width=.4\textwidth]{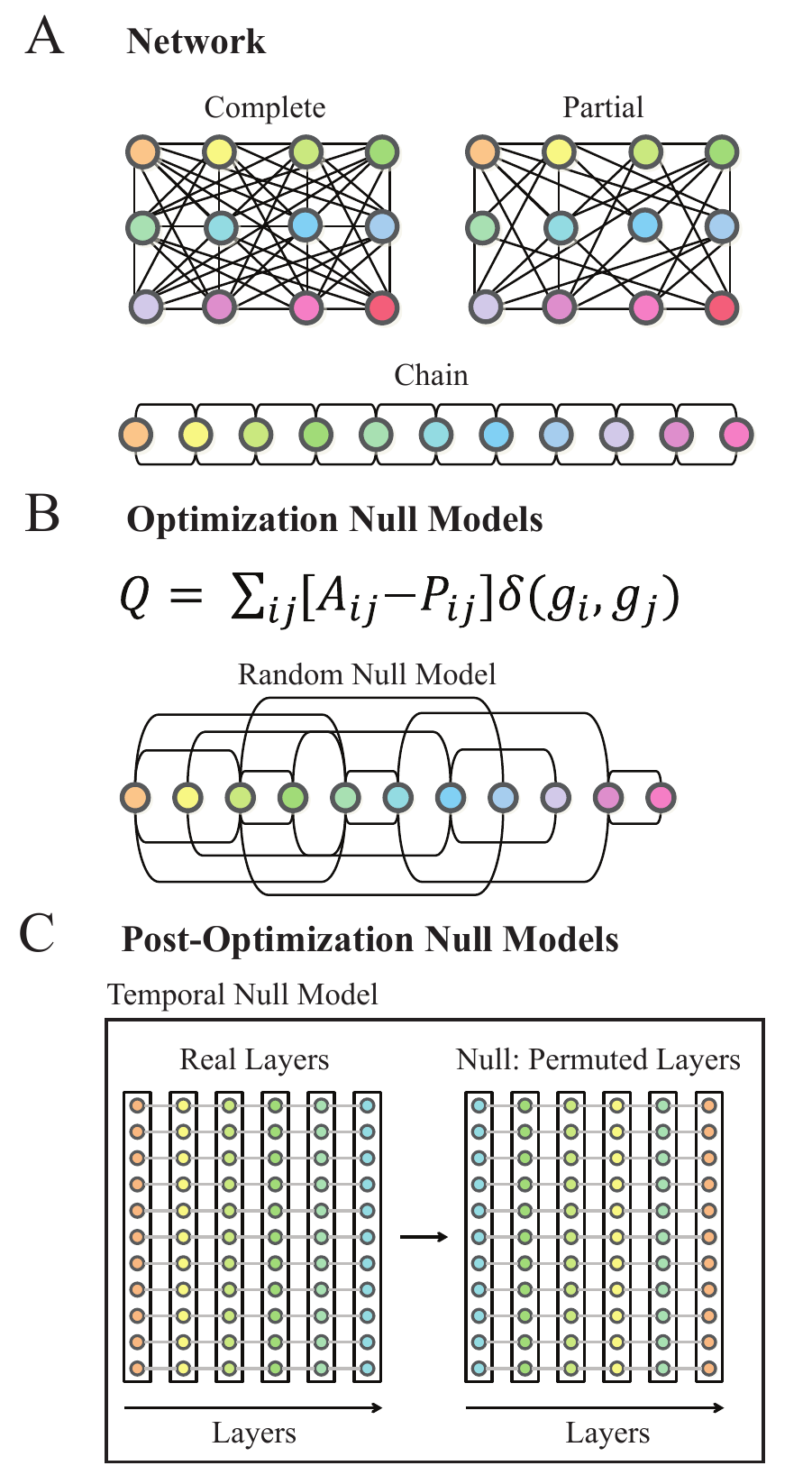}
\caption[]{(Color Online) Methodological considerations important in the
investigation of dynamic community structure in temporal networks. (\emph{A})
Depending on the system under study, a single network layer (which is
represented using an ordinary adjacency matrix with an extra index to
indicate the layer) might by definition only allow edges from some subset of
the complete set of node pairs, as is the case in the depicted chain-like
graph. We call such a situation \emph{partial connectivity}. (\emph{B})
Although the most common optimization null model employs random graphs (e.g.,
the Newman-Girvan null model, which is closely related to the configuration
model \cite{Porter2009,Newman2010}), other models can also provide important
insights into network community structure. (\emph{C}) After determining a set
of partitions that maximize the modularity $Q$ (or a similar quality
function), it is interesting to test whether the community structure is
different from, for example, what would be expected with a scrambling of time
layers (i.e., a temporal null model) or node identities (i.e., a nodal null
model) \cite{Bassett2011b}. \label{Fig1}}
\end{figure}

In the present paper, we examine null models in time-dependent networks and
investigate their use in the algorithmic detection of cohesive, dynamic
communities in such networks (see Fig.~\ref{Fig1}).  Indeed, community
detection in temporal networks necessitates the development of null models
that are appropriate for such networks. Such null models can help provide
bases of comparison at various stages of the community-detection process, and
they can thereby facilitate the principled identification of dynamic
structure in networks. Indeed, the importance of developing null models
extends beyond community detection, as such models make it possible to obtain
statistically significant estimates of network diagnostics.

Our dynamic network null models fall into two categories: \emph{optimization}
null models, which we use in the identification of community structure; and
\emph{post-optimization} null models, which we use to examine the identified
community structure.  We describe how these null models can be selected in a
manner appropriate to known features of a network's construction, identify
potentially interesting network scales by determining values of interest for
structural and temporal resolution parameters, and inform the choice of
representative partitions of a network into communities.

%%%%%%%%%%

\section{Methods}

\subsection{Community Detection}

Community-detection algorithms provide ways to decompose a network into dense
groups of nodes called `modules' or `communities'. Intuitively, a community
consists of a set of nodes that are connected among one another more densely
than they are to nodes in other communities.  A popular way to identify
community structure is to optimize a \emph{quality function}, which can be
used to measure the relative densities of intra-community connections versus
inter-community connections.  See \cite{Porter2009,Fortunato2010,Newman2012}
for recent reviews on network community structure and
\cite{Good2010,Fortunato2007,Bickel2009,Hutchings2011} for
discussions of various caveats that should be considered when optimizing
quality functions to detect communities.

One begins with a network of $N$ nodes and a given set of connections between
those nodes. In the usual case of single-layer networks (e.g., static
networks with only one type of edge), one represents a network using an
$N\times N$ adjacency matrix $\mathbf{A}$.  The element $A_{ij}$ of the adjacency
matrix indicates a direct connection or `edge' from node $i$ to node $j$, and
its value indicates the weight of that connection.  The quality of a hard
partition of $\mathbf{A}$ into communities (whereby each node is assigned to
exactly one community) can be quantified using a quality function.  The
most popular choice is  \emph{modularity} \cite{NG2004,markfast,Newman2006,Porter2009,Fortunato2010}
\begin{equation}\label{one}
	Q_{0} = \sum_{ij} [A_{ij} - \gamma P_{ij}] \delta(g_{i},g_{j})\,,
\end{equation}
where node $i$ is assigned to community $g_{i}$, node $j$ is assigned to
community $g_{j}$, the Kronecker delta $\delta(g_{i},g_{j})=1$ if $g_{i} =
g_{j}$ and it equals $0$ otherwise, $\gamma$ is a resolution parameter (which
we will call a \emph{structural resolution parameter}), and $P_{ij}$ is the
expected weight of the edge connecting node $i$ to node $j$ under a specified
null model. The choice $\gamma  = 1$ is very common, but it is important to
consider multiple values of $\gamma$ to examine groups at multiple scales
\cite{rb2006,Porter2009,Onnela2011}. Maximization of $Q_{0}$ yields a hard
partition of a network into communities such that the total edge weight
inside of modules is as large as possible (relative to the null model and
subject to the limitations of the employed computational heuristics, as
optimizing $Q_{0}$ is NP-hard \cite{Porter2009,Fortunato2010,Brandes2008}).

Recently, the null model in the quality function (\ref{one}) has been
generalized so that one can consider sets of $L$ adjacency matrices, which
are combined to form a rank-3 adjacency tensor ${\bf A}$ that can be used to
represent time-dependent or multiplex networks.  One can thereby define a
\emph{multilayer modularity} (also called `multislice modularity') \cite{Mucha2010}
\begin{equation} \label{eq:Qml}
    	Q = \frac{1}{2\mu}\sum_{ijlr}\left\{\left(A_{ijl}-\gamma_l P_{ijl} \right)\delta_{lr} + \delta_{ij}\omega_{jlr}\right\} \delta(g_{il},g_{jr})\,,
\end{equation}
where the adjacency matrix of layer $l$ has components $A_{ijl}$, the element
$P_{ijl}$ gives the components of the corresponding layer-$l$ matrix for the
optimization null model, $\gamma_l$ is the structural resolution parameter of
layer $l$, the quantity $g_{il}$ gives the community assignment of node $i$ in layer $l$,
the quantity $g_{jr}$ gives the community assignment of node $j$ in layer $r$,
the element $\omega_{jlr}$ gives the connection strength (i.e., an `interlayer coupling
parameter', which one can call a \emph{temporal resolution parameter} if one
is using the adjacency tensor to represent a time-dependent network) from
node $j$ in layer $r$ to node $j$ in layer $l$, the total edge weight in the
network is $\mu=\frac{1}{2}\sum_{jr} \kappa_{jr}$, the strength (i.e., weighted degree) of node $j$
in layer $l$ is $\kappa_{jl}=k_{jl}+c_{jl}$, the intra-layer strength of node
$j$ in layer $l$ is $k_{jl}=\sum_i A_{ill}$, and the inter-layer strength of node $j$ in
layer $l$ is $c_{jl} = \sum_r \omega_{jlr}$.

Equivalent representations that use other notation can, of
course, be useful.  For example, multilayer modularity can be recast as a set
of rank-2 matrices describing connections between the set of all nodes across
layers [e.g., for spectral partitioning
\cite{Newman2006,Newman2006b,Richardson2009}].
One can similarly generalize $Q$ for higher-rank tensors, which one would use when
studying community structure in networks that are both time-dependent and
multiplex, through appropriate specification of inter-layer coupling tensors.

%%%%%%%%%%

\subsection{Network Diagnostics}

To characterize multilayer community structure, we compute four example
diagnostics for each hard partition: the modularity $Q$, the number of modules $n$, the mean community size $s$ (which is equal to the number of nodes in the community and is
proportional to $1/n$), and the stationarity $\zeta$ \cite{Palla2007}. To
compute $\zeta$, we calculate the autocorrelation function $U(t,t+m)$ of
two states of the same community $G(t)$ at $m$ time steps (i.e., $m$ network
layers) apart:
\begin{equation}
	U(t,t+m) \equiv \frac{|G(t)\cap G(t+m)|}{|G(t)\cup G(t+m)|}\,,
\end{equation}
where $|G(t)\cap G(t+m)|$ is the number of nodes that are members of both
$G(t)$ and $G(t+m)$, and $|G(t)\cup G(t+m)|$ is the number of nodes in the union of the community at times $t$ and $t+m$. Defining $t_{0}$ to be the first time step in which the
community exists and $t'$ to be the last time in which it exists, the
stationarity of a community is \cite{Palla2007}
\begin{equation}\label{zeta}
	\zeta \equiv \frac{\sum_{t=t_{0}}^{t'-1} U(t,t+1)}{t'-t_{0}}\,.
\end{equation}
This gives the mean autocorrelation over consecutive time steps
\footnote{Equation (\ref{zeta}) gives the definition for the notion of
stationarity that we used in \cite{Bassett2011b}.  The equation for this
quantity in \cite{Bassett2011b} has a typo in the denominator. We wrote
incorrectly in that paper that the denominator was $t'-t_{0} - 1$, whereas
the numerical computations in that paper used $t'-t_{0}$.}.

In addition to these diagnostics, which are defined using the entire multilayer community
structure, we also compute two example diagnostics on the community structures
of the component layers: the mean single-layer modularity $\langle Q_{s}
\rangle$ and the variance $\mathrm{var}(Q_{s})$ of the single-layer
modularity over all layers. The single-layer modularity $Q_{s}$ is defined as
the static modularity quality function, $Q_{s} = \sum_{ij} [A_{ij} - \gamma
P_{ij}] \delta(g_{i},g_{j})$, computed for the partition $g$ that we obtained via
optimization of the multilayer modularity function $Q$. We have chosen to
use a few simple ways to help characterize the time series for $Q_{s}$, though of
course other diagnostics should also be informative.

%%%%%%%%%%%%%

\section{
%Temporal Network
Data Sets}

We illustrate the concept and uses of dynamic network null models using two
example network ensembles: (1) 75-time-layer brain networks drawn from each of 20
human subjects and (2) behavioral networks with about 150 time layers drawn from each
of 22 human subjects. Importantly, the use of network ensembles makes it possible
to examine robust structure (and also its variance) over multiple network
instantiations. We have previously examined both data sets in the context of
neuroscientific questions \cite{Bassett2011b,Wymbs2011}. In this paper, we
use them as illustrative examples for the consideration of methodological
issues in the detection of dynamic communities in temporal networks.

These two data sets, which provide examples of different types of network data,
illustrate a variety of issues in network construction: (1) node and edge
definitions, (2) complete versus partial connectivity, (3) ordered versus
categorical nodes, and (4) confidence in edge weights. In many fields, determining the
definition of nodes and edges is itself an active area of investigation
\footnote{In other areas of investigation, it probably should be!}.  See, for example, several recent papers that address such questions in the context of large-scale human brain networks
\cite{Bullmore2011,Bassett2010c,Zalesky2010,Wang2009,Bialonski2010,Ioannides2007}
and in networks more generally \cite{Butts2009}.  Another important issue is
whether to examine a given adjacency matrix in an exploratory manner or to
impose structure on it based on \emph{a priori} knowledge. For example, when
nodes are \emph{categorical}, one might represent their relations using a fully connected network and then identify communities of any group of nodes. However, when
nodes are \emph{ordered} --- and particularly when they are in a chain of
weighted nearest-neighbor connections --- one expects communities to
group neighboring nodes in sequence, as typical community-detection methods are unlikely to yield many out-of-sequence jumps in community assignment. The issue of confidence in the estimation of edge weights is also very important, as it can prompt an investigator to delete edges from a network when their statistical validity is questionable. A closely related
issue is how to deal with known or expected missing data, which can affect
either the presence or absence of nodes themselves or the weights of edges
\cite{Kossinets2006,Clauset2008,Guimera2009,Kim2011}.

%%%%%%%

\subsection{Data Set 1: Brain Networks}

Our first data set contains categorical nodes with partial connectivity and
variable confidence in edge weights. The nodes remain unchanged in time, and
edge weights are based on covariance of node properties. This covariance
structure is non-local in the sense that weights exist between both
topologically neighboring nodes and topologically distant nodes
\cite{Sporns2010,Bullmore2009}. This property has been linked in other
dynamical systems to behaviors such as chimera states, which coherent and
incoherent regions coexist \cite{Abrams2004,Kuramoto2002,Shima2004}. Another
interesting feature of this data set is that it is drawn from an experimental
measurement with high spatial resolution (on the order of centimeters) but
relatively poor temporal resolution (on the order of seconds).

As described in more detail in Ref.~\cite{Bassett2011b}, we construct an
ensemble of networks (20 individuals over 3 experiments, which yields 60 multilayer
networks) that represent the functional connectivity between large regions of
the human brain. In these networks, $N=112$
centimeter-scale, anatomically distinct brain regions are our (categorical)
network nodes. We study the temporal interaction of these nodes --- such interactions are thought to underly cognitive function --- by first measuring their activity every 2
seconds during simple finger movements using functional magnetic resonance
imaging (fMRI). We cut these regional time series into time slices (which yield layers in the multilayer network) of roughly 3-minute duration. Each such
layer corresponds to a time series whose length is 80 units.

To estimate the interactions (i.e., edge weights) between nodes, we calculate a measure
of statistical similarity between regional activity profiles
\cite{Friston1994}. Using a wavelet transform, we extract
frequency-specific activity from each time series in the range 0.06--0.12 Hz.
For each time layer $l$ and each pair of regions $i$ and $j$, we define the
weight of an edge connecting region $i$ to region $j$ using the coherence
between the wavelet-coefficient time series in each region, and these weights
form the elements of a weighted, undirected temporal network ${\bf W}$ with components
$W_{ijl} = W_{jil}$. The \emph{magnitude-squared coherence} $G_{ij}$ between time series $i$ and $j$ is a function of frequency.  It is defined by the equation
\begin{equation}
	G_{ij}(f)=\frac{|F_{ij}(f)|^2}{F_{ii}(f)F_{jj}(f)}\,,
\end{equation}
where $F_{ii}(f)$ and $F_{jj}(f)$ are the power spectral density functions of
$i$ and $j$, respectively, and $F_{ij}(f)$ is the cross-power spectral
density function of $i$ and $j$. We let $H_{ij}$ denote the mean of
$G_{ij}(f)$ over the frequency band of interest, and the weight of edge
$W_{ijl}$ is equal to $H_{ij}$ computed for layer $l$.

We use a false-discovery rate
\cite{Benjamini2001} to threshold connections whose coherence values are not significantly greater than that expected at random. This yields a multilayer network ${\bf A}$
with components $A_{ijl}$ (i.e., a rank-3 adjacency tensor). The nonzero entries in
$A_{ijl}$ retain their weights.  We couple the layers of $A_{ijl}$ to one
another with temporal resolution parameters of weight $\omega_{jlr}$ between node $j$ in layer $r$ and node $j$ in layer $l$. In this
paper, we let $\omega_{jlr}\equiv\omega \in [0.1,40]$ be identical between each node $j$ in a given layer with itself in nearest-neighbor layers.  (In all other cases, $\omega_{jlr} = 0$.)

In Fig.~\ref{Fig2}A, we show an example time layer from $A_{ijl}$ for a
single subject in this experimental data. In this example, the statistical
threshold is evinced by the set of matrix elements set to $0$. Because brain
network nodes are categorical, one can apply community detection algorithms
in these situations to identify communities composed of any set of nodes.
(Note that the same node from different times can be assigned to the same
community even if the node is assigned to other communities at intervening
times.) One biological interpretation of network communities in brain
networks is that they represent groups of nodes that serve distinct cognitive
functions (e.g., vision, memory, etc.) that can vary in time
\cite{Bullmore2012,Doron2012}.

%%%%%%%%%%

\subsection{Data Set 2: Behavioral Networks}

Our second data set contains ordered nodes that remain unchanged in time. The
network topology in this case is highly constrained, as edges are only
present between consecutive nodes. (We call this `nearest-neighbor'
coupling.) Another interesting feature of this data set is that the number of
time slices is an order of magnitude larger than the number of nodes in a
slice.

As described in more detail in Ref.~\cite{Wymbs2011}, we construct an ensemble
of 66 behavioral networks from 22 individuals and 3 experimental conditions.
These networks represent a set of finger movements in the same simple motor
learning experiment from which we constructed the brain networks in data set
1. Subjects were instructed to press a sequence of buttons corresponding to a
sequence of $12$ pseudo-musical notes shown to them on a screen.

Each node represents an interval between consecutive button presses.  A single network layer consists of $N = 11$ nodes (i.e., there is one interval between each pair of notes), which are
connected in a chain via weighted, undirected edges.  In
Ref.~\cite{Wymbs2011}, we examined the phenomenon of motor `chunking', which is a
fascinating but poorly-understood phenomenon in which groups of movements are
made with similar inter-movement durations. (This is similar to remembering a phone
number in groups of a few digits or grouping notes together as one masters how to play a song.)  For each experimental trial $l$ and each pair of inter-movement intervals $i$ and $j$, we define the weight of an edge connecting inter-movement $i$ to inter-movement $j$ as the normalized
similarity in inter-movement durations. The \emph{normalized similarity} between
nodes $i$ and $j$ is defined as
\begin{equation}
	\rho_{ijl} = \frac{\bar{d}_{l} - d_{ijl}}{\bar{d}_{l}}\,,
\end{equation}
where $d_{ijl}$ is the absolute value of the difference of lengths of the
$i$th and $j$th inter-movement time intervals in trial $l$ and $\bar{d}_{l}$ is the maximum
value of $d_{ijl}$ in trial $l$. These weights yield the elements $W_{ijl}$
of a weighted, undirected multilayer network ${\bf W}$. Because finger movements occur in series,
inter-movement $i$ is connected in time to inter-movement $i \pm 1$ but not to
any other inter-movements $i+n$ for $|n| \neq 1$.

\begin{figure}[]
\includegraphics[width=.4\textwidth]{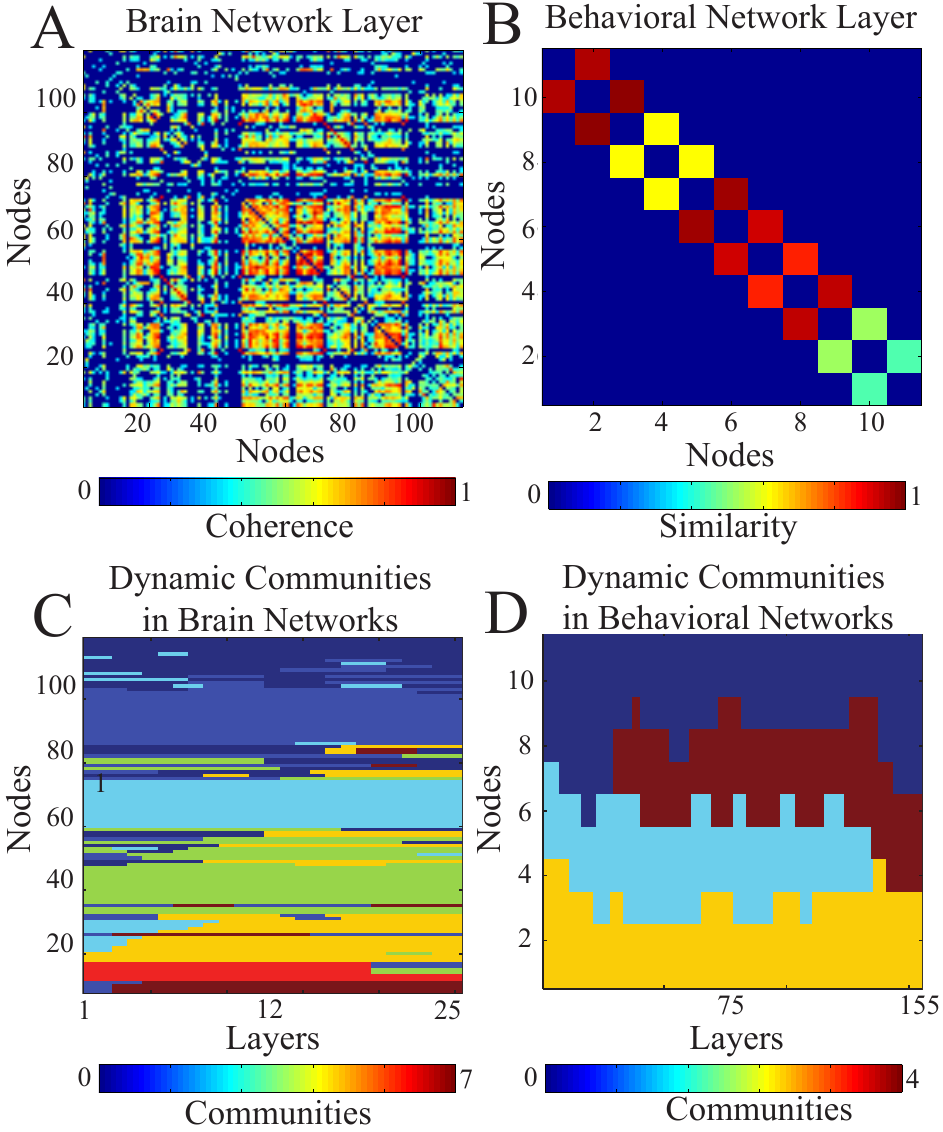}
\caption[]{(Color Online) Network layers and community assignments from two
example data sets: (\emph{A}) a brain network based on correlations between
blood-oxygen-level-dependent (BOLD) signals \cite{Bassett2011b} and (\emph{B}) a
behavioral network based on similarities in movement times during
a simple motor learning experiment \cite{Wymbs2011}.  We use these data
sets to illustrate situations with categorical
nodes and ordered nodes, respectively. In the bottom panels, we show community assignments obtained using multilayer community detection for (\emph{C}) the brain networks and (\emph{D}) the behavioral networks.
\label{Fig2}}
\end{figure}

To encode this conceptual relationship as a network, we set all
non-contiguous connections in $W_{ijl}$ to $0$ and thereby construct a
weighted, undirected \emph{chain network} $A_{ijl}$. In Fig.~\ref{Fig2}B, we show an example
trial layer from $A_{ijl}$ for a single subject in this experimental data.
We couple layers of $A_{ijl}$ to one another with weight $\omega_{jlr}$,
which gives the connection strength between node $j$ in experimental trial $r$
and node $j$ in trial $l$. In a given instantiation of the network, we again let
$\omega_{jlr} \equiv \omega \in [0.1,40]$ be identical for all nodes $j$ for all connections between nearest-neighbor layers.  (Again, $\omega_{jlr} = 0$ in all other cases.) Because inter-movement
nodes are ordered, one can apply community-detection algorithms to identify
communities of nodes in sequence.  Each community represents a motor `chunk'.

%%%%%%%%%%%

\section{Results}

\subsection{Modularity-Optimization Null Models}

After constructing a multilayer network ${\bf A}$ with elements $A_{ijl}$, it is necessary to select
an optimization null model $P_{ijl}$ in equation \eqref{eq:Qml}. The most
common modularity-optimization null model used in undirected, single-layer networks is the Newman-Girvan null model
\cite{markfast,NG2004,Newman2006,Porter2009,Fortunato2010}
\begin{equation}\label{pij}
	P_{ij} = \frac{k_{i} k_{j}}{2m}\,,
\end{equation}
where $k_{i}=\sum_j A_{ij}$ is the strength of node $i$ and $m=\frac{1}{2}\sum_{ij} A_{ij}$. The definition (\ref{pij}) can be
extended to multilayer networks using
\begin{equation}\label{pijl}
	P_{ijl} = \frac{k_{il}k_{jl}}{2m_l}\,,
\end{equation}
where $k_{il}=\sum_j A_{ijl}$ is the strength of node $i$ in layer $l$ and $m_l=\frac{1}{2}\sum_{ij} A_{ijl}$. Optimization of $Q$ using the null model (\ref{pijl}) identifies partitions
of a network into groups that have more connections (in the case of binary
networks) or higher connection densities (in the case of weighted networks)
than would be expected for the distribution of connections (or
connection densities) expected in a null model. We use the notation
$\mathbf{A}^{l}$ for the layer-$l$ adjacency matrix composed of elements
$A_{ijl}$ and the notation $\mathbf{P}^{l}$ to denote the layer-$l$ null-model matrix
with elements $P_{ijl}$. See Fig.~\ref{Fig3}A for an example layer
$\mathbf{A}^{l}$ from a multilayer behavioral network and Fig.~\ref{Fig3}B
for an example instantiation of the Newman-Girvan null model
$\mathbf{P}^{l}$.

%%%%%%%%%%%

\subsubsection{Optimization Null Models for Ordered Node Networks}

\begin{figure*}[]
\includegraphics[width=.6\textwidth]{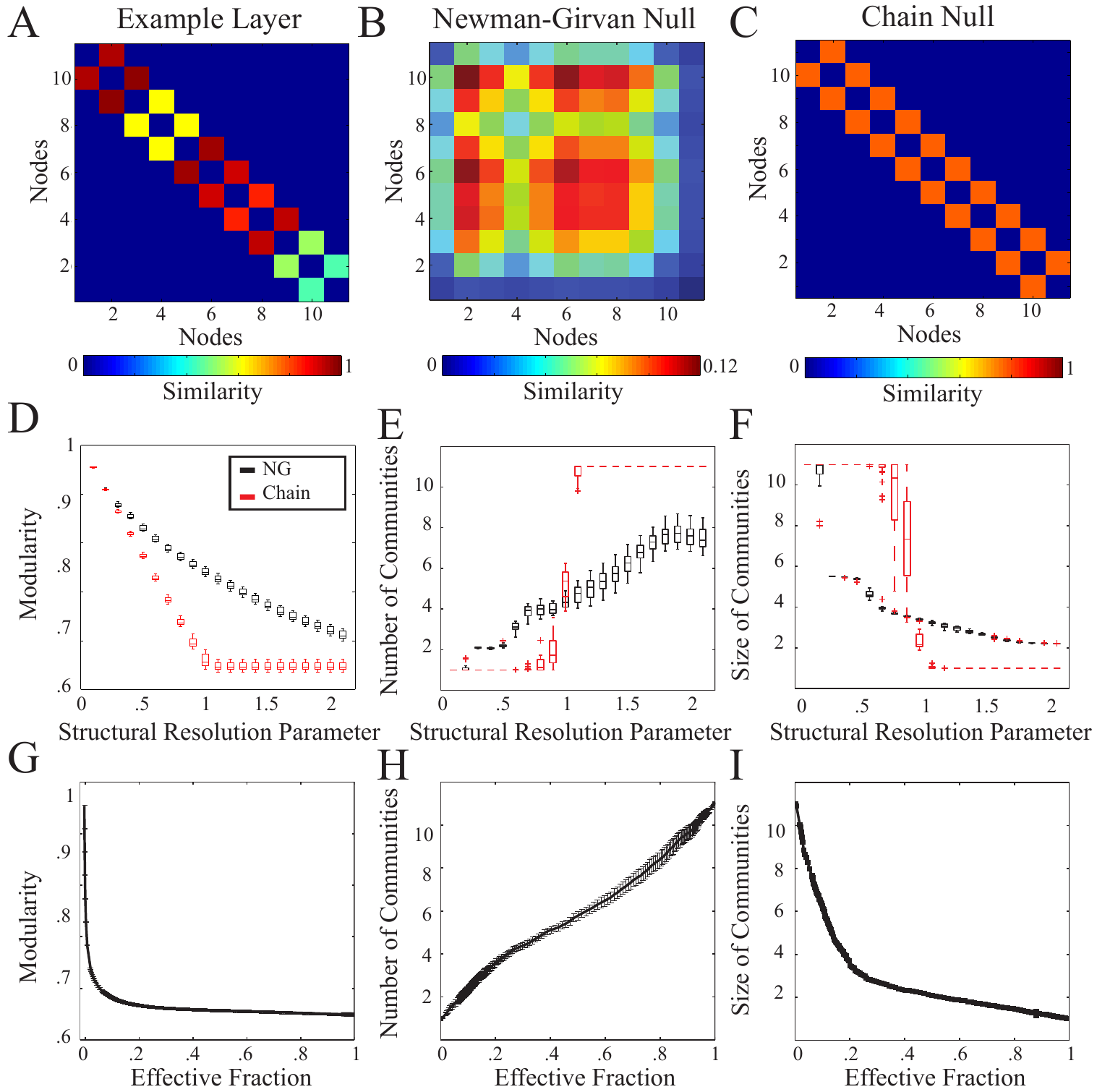}
\caption[]{(Color Online) Modularity-optimization null models. (\emph{A})
Example layer $\mathbf{A}^{l}$ from a behavioral network. (\emph{B})
Newman-Girvan and (\emph{C}) chain null models $\mathbf{P}^{l}$ for the layer
shown in panel (\emph{A}). (\emph{D}) Optimized multilayer modularity value
$Q$, (\emph{E}) number of communities $n$, and (\emph{F}) mean community size
$s$ for the complete multilayer behavioral network employing the
Newman-Girvan (black) and chain (red) optimization null models as a function
of the structural resolution parameter $\gamma$. (\emph{G}) Optimized
modularity value $Q$, (\emph{H}) number of communities $n$, and (\emph{I})
mean community size $s$ for the multilayer behavioral network employing chain
optimization null models as a function of the effective fraction
$\xi_{\mathrm{ml}}(\gamma)$ of edges that have larger weights than their
null-model counterparts. We averaged the values of $Q$, $n$, and $s$ over the
3 different 12-note sequences and $C = 100$ optimizations. Box plots in
(\emph{D}-\emph{F}) indicate quartiles and 95\% confidence intervals over the
22 individuals in the study. The error bars in panels (\emph{G}-\emph{I})
indicate a standard deviation from the mean.  In some instances, this is
smaller than the line width. The temporal resolution-parameter value is
$\omega = 1$. \label{Fig3}}
\end{figure*}

The Newman-Girvan null model is particularly useful for networks with categorical nodes, in which
a connection between any pair of nodes can occur in theory. However,
when using a chain network of ordered nodes, it is
useful to consider alternative null models. For example, in an ordinary network (i.e., one that is represented using an adjacency matrix), one can define
\begin{equation}
	P_{ij} = \rho A'_{ij}\,,
\end{equation}
where $\rho$ is the mean edge weight of the chain network and $A'_{ij}$ is
the binarized version of $A_{ij}$, in which nonzero elements of $A_{ij}$ are
set to $1$ and zero-valued elements remain unaltered.  Such a null model can
also be defined for multilayer networks:
\begin{equation}
	P_{ijl} = \rho_{l} A'_{ijl}\,,
\end{equation}
where $\rho_{l}$ is the mean edge weight in layer $l$ and
$A'_{ijl}$ is the binarized version of $A_{ijl}$. The optimization of $Q$
using this null model identifies partitions of a network whose communities
have a larger strength than the mean. See Fig.~\ref{Fig3}C for an example
of this chain null model $\mathbf{P}^{l}$ for the behavioral
network layer shown in Fig.~\ref{Fig3}A.

In Fig.~\ref{Fig3}D, we illustrate the effect that the choice of optimization
null model has on the modularity values $Q$ of the behavioral networks as a
function of the structural resolution parameter. (Throughout the manuscript, we use a Louvain-like locally-greedy algorithm to maximize the multilayer modularity quality function
\cite{Blondel2008,Jutla2012}.) The Newman-Girvan null model gives decreasing
values of $Q$ for $\gamma \in [0.1,2.1]$, whereas the chain null model
produces lower values of $Q$, which behaves in a qualitatively different manner for
$\gamma < 1$ versus $\gamma > 1$.  To help understand this feature, we plot the number and
mean size of communities as a function of $\gamma$ in
Figs.~\ref{Fig3}E and \ref{Fig3}F.  As $\gamma$ is increased, the Newman-Girvan null model
yields network partitions that contain progressively more communities (with
progressively smaller mean size).  The number of communities that we obtain
in partitions using the chain null model also increases with $\gamma$, but it
does so less gradually. For $\gamma \ll 1$, one obtains a network partition
consisting of a single community of size $N_{l}=11$; for $\gamma \gg 1$, each
node is instead placed in its own community.  For $\gamma=1$, nodes are
assigned to several communities whose constituents vary with time (see, for
example, Fig.~\ref{Fig2}D).

The above results highlight the sensitivity of network diagnostics such as
$Q$, $n$, and $s$ to the choice of an optimization null model.  It is important
to consider this type of sensitivity in the light of other known issues, such
as the extreme near-degeneracy of quality functions like modularity
\cite{Good2010}. Importantly, the use of the chain null models provides a
clear delineation of network behavior in this example into three regimes
as a function of $\gamma$: a single community with variable $Q$ (low $\gamma$),
a variable number of communities as $Q$ reaches a minimum value ($\gamma
\approx 1$), and a set of singleton communities with minimum $Q$ (high
$\gamma$). This illustrates that it is crucial to consider a null model
appropriate for a given network, as it can provide more interpretable results
than just using the usual choices (such as the Newman-Girvan null model).

The structural resolution parameter $\gamma$ can be transformed so that it
measures the effective fraction of edges $\xi(\gamma)$ that have larger
weights than their null-model counterparts \cite{Onnela2011}. One can define
a generalization of $\xi$ to multilayer networks, which allows one to examine
the behavior of the chain null model near $\gamma=1$ in more detail. For each
layer $l$, we define a matrix $\mathbf{X}^{l}(\gamma)$ with elements
$X_{ijl}(\gamma) = A_{ijl}- \gamma P_{ijl}$, and we then define
$c^{X}(\gamma)$ to be the number of elements of $\mathbf{X}^{l}(\gamma)$ that
are less than $0$. We sum $c^{X}(\gamma)$ over layers in the multilayer
network to construct $c^{X}_{\mathrm{ml}}(\gamma)$. The transformed
structural resolution parameter is then given by
\begin{equation}
	\xi_{\mathrm{ml}}(\gamma) = \frac{c^{X}_{\mathrm{ml}}(\gamma) - c^{X}_{\mathrm{ml}}(\Lambda_{\mathrm{min}})}{c^{X}_{\mathrm{ml}}(\Lambda_{\mathrm{max}}) - c^{X}_{\mathrm{ml}}(\Lambda_{\mathrm{min}})}\,,
\end{equation}
where $\Lambda_{\mathrm{min}}$ is the value of $\gamma$ for which the network
still forms a single community in the multilayer optimization and
$\Lambda_{\mathrm{max}}$ is the value of $\gamma$ for which the network still
forms $N$ singleton communities in the multilayer optimization.  (We use
Roman typeface in the subscripts in $c^{X}_{\mathrm{ml}}$ and
$\xi_{\mathrm{ml}}$ to emphasize that we are describing multilayer objects
and, in particular, that the subscripts do not represent indices.) In
Figs.~\ref{Fig3}G-I, we report the optimized (i.e., maximized) modularity
value, the number of communities, and the mean community size as functions of
the transformed structural resolution parameter $\xi_{\mathrm{ml}}(\gamma)$.
(Compare these plots to Figs.~\ref{Fig3}D-F.) For all three diagnostics, the
apparent transition points seem to be more gradual as a function of
$\xi_{\mathrm{ml}}(\gamma)$ than they are as a function of $\gamma$.  For
systems like the present one that do not exhibit a pronounced, nontrivial
plateau in these diagnostics as a function of a structural resolution
parameter, it might be helpful to have \emph{a priori} knowledge about the
expected number or sizes of communities (see, e.g., \cite{Wymbs2011}) to help
guide further investigation.

%%%%%%%%%%

\subsubsection{Optimization Null Models for Networks Derived from Time Series}

\begin{figure*}[]
\includegraphics[width=.75\textwidth]{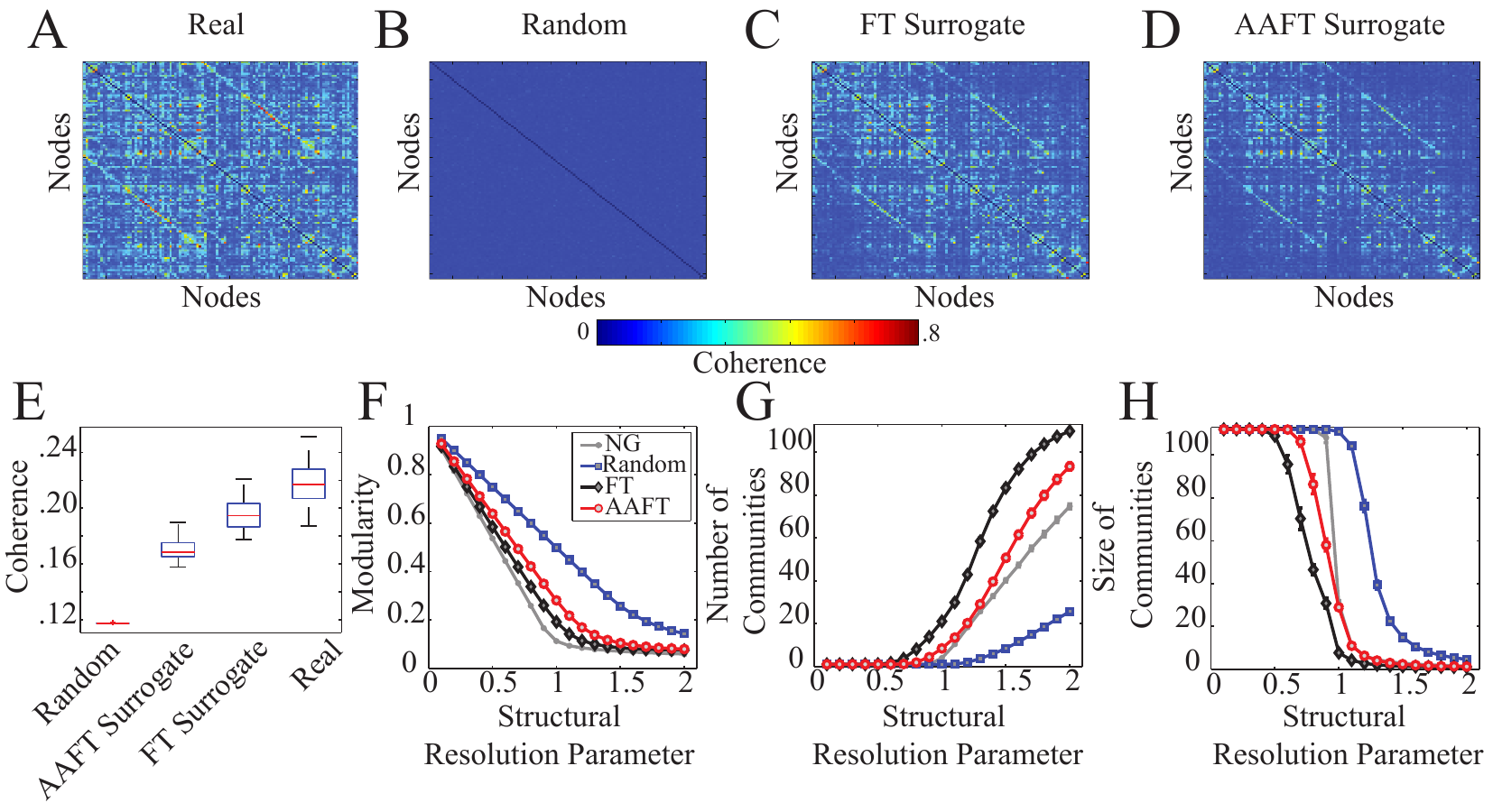}
\caption[]{(Color Online) Modularity-optimization null models for time
series. (\emph{A}) Example coherence matrix $\mathbf{A}^{l}$ averaged over
layers from a brain network. (\emph{B}) Random time shuffle, (\emph{C})
Fourier transform (FT) surrogate, and (\emph{D}) amplitude-adjusted Fourier
transform (AAFT) surrogate null models $\mathbf{P}^{l}$ averaged over layers.
(\emph{E}) Coherence of each matrix type averaged over subjects, scans, and
layers. We note that the apparent lack of structure in (\emph{B}) is
partially related to its significantly decreased coherence in comparison to
the other models. (\emph{F}) Optimized modularity values $Q$, (\emph{G}) number of
communities $n$, and (\emph{H}) mean community size $s$ for the
multilayer brain network employing the Newman-Girvan (black), random
time-shuffle (blue), FT surrogate (gray), and AAFT surrogate (red) optimization null models
as functions of the structural resolution parameter $\gamma$. We averaged
the values of these diagnostics over 3 different scanning sessions and $C
= 100$ optimizations. Box plots indicate quartiles and 95\% confidence intervals over the
20 individuals in the study. The temporal resolution parameter is $\omega =
1$. \label{Fig34}}
\end{figure*}

Although the Newman-Girvan null model can be used in networks with
categorical nodes, such as the brain networks in data set 1 (see
Fig.~\ref{Fig34}A), it does not take advantage of the fact that these
networks are derived from similarities in time series. Accordingly, we
generate surrogate data to construct two dynamic network null models for
community detection that might be particularly appropriate for networks
derived from time-series data.

First, we note that a simple null model (which we call `Random') for time
series is to randomize the elements of the time-series vector for each node
before computing the similarity matrix (see Fig.~\ref{Fig34}B) \footnote{A
discrete time series can be represented as a vector.  A continuous time
series would first need to be discretized.}. However, the resulting time
series do not have the mean or variance of the original time series, and this
yields a correlation- or coherence-based network with very low edge weights.
To preserve the mean, variance, and autocorrelation function of the original
time series, we employ a surrogate-data generation method that scrambles the
phase of time series in Fourier space \cite{Prichard1994}. Specifically, we
assume that the linear properties of the time series are specified by the
squared amplitudes of the discrete Fourier transform
\begin{equation}
	|S(u)|^2 = \left| \frac{1}{\sqrt{V}} \sum_{v=0}^{V-1} s_{v} e^{i 2 \pi u v / V}\right|^{2}\,,
\end{equation}
where $s_{v}$ denotes an element in a time series of length $V$. (That is, $V$ is the number of elements in the time-series vector.)  We construct surrogate data by multiplying the Fourier transform by phases chosen uniformly at random and transforming back to the time domain:
\begin{equation}
	\bar{s}_{v} = \frac{1}{\sqrt{V}} \sum_{v=0}^{V-1} e^{i a_{u}} |S_{u}| e^{i 2 \pi k v / V}\,,
\end{equation}
where $a_{u}\in [0,2 \pi)$ are chosen independently and uniformly at random.
\footnote{The code used for this computation actually operates on $[0,2
\pi]$. However, this should be an equivalent mathematical estimate to the
same computation on $[0,2 \pi)$, which is the same except for a set of
measure zero.} This method, which we call the \emph{Fourier transform (FT)
surrogate} (see Fig.~\ref{Fig34}C), has been used previously to construct
covariance matrices \cite{Nakatani2003} and to characterize networks
\cite{Zalesky2012}. A modification of this method, which we call the
\emph{amplitude-adjusted Fourier transform (AAFT) surrogate}, allows one to
also retain the amplitude distribution of the original signal
\cite{Theiler1992} (see Fig.~\ref{Fig34}D). One can alter nonlinear
relationships between time series while preserving linear relationships
between time series by applying an identical shuffling to both time series;
one can alter both linear and nonlinear relationships between time series by
applying independent shufflings to each time series \cite{Prichard1994}.

We demonstrate in Fig.~\ref{Fig34}E that, among the four null models that we
consider, the mean coherence of pairs of FT surrogate series match most
closely to that of the original data.  Pairs of Random time series have the
smallest mean coherence, and pairs of AAFT surrogate series have the next
smallest.  The fact that the AAFT surrogate is less like the real data (in
terms of mean coherence) than the simpler FT surrogate might stem from a
rescaling step \cite{Nakatani2003} that causes the power spectrum to whiten
(i.e., the step flattens the power spectral density) \footnote{It is also
important to note that the AAFT method can allow nonlinear correlations to
remain in the surrogate data. Therefore the development of alternative
surrogate data generation methods might be necessary
\cite{surrogate1,surrogate2}.}. In Figs.~\ref{Fig34}F-H, we show three
diagnostics (optimized modularity, mean community size, and number of
communities) as a function of the structural resolution parameter $\gamma$
for the various optimization null models. We note that the Newman-Girvan null
model produces the smallest $Q$ value and a middling community size, whereas
the surrogate time series models produce higher $Q$ values and more
communities of smaller mean size. The Random null model produces the largest
value of $Q$ and the fewest communities, which is consistent with the fact
that it contains the smallest amount of shared information (i.e., mean
coherence) with the real network.

%%%%%%%%%%%%%%

\subsection{Post-Optimization Null Models}

After identifying the partition(s) that maximize modularity, one might wish
to determine whether the identified community structure is significantly
different from that expected under other null hypotheses. For example, one
might wish to know whether any temporal evolution is evident in the dynamic
community structure (see Fig.~\ref{Fig1}C). To do this, one can employ
\emph{post-optimization null models}, in which a multilayer network is
scrambled in some way to produce a new multilayer network. One can then
maximize the modularity of the new network and compare the resulting
community structure to that obtained using the original network.
Unsurprisingly, one's choice of post-optimization null model should be
influenced by the question of interest, and it should also be constrained by
properties of the network under examination. We explore such influences and
constraints using our example networks.

%%%%%%%%%%%%%

\subsubsection{Intra-Layer and Inter-Layer Null Models}

There are various ways to construct \emph{connectional null models} (i.e.,
intra-layer null models), which randomize the connectivity structure within a
network layer ($\mathbf{A}^{l}$) \cite{Bassett2011b} \footnote{In the
descriptions below, we use terms like `random' rewiring to refer to a process
that we are applying uniformly at random aside from specified constraints.}.
For binary networks, one can obtain ensembles of random graphs with the same
mean degree as that of a real network using Erd\H{o}s-R\'{e}nyi random graphs
\cite{Newman2010}, and ensembles of weighted random networks can similarly be
constructed from weighted random graph models \cite{Garlaschelli2009}. To
retain both the mean and distribution of edge weights, one can employ a
permutation-based connectional null model that randomly rewires network edges
with no additional constraints by reassigning uniformly at random the entire
set of matrix elements $A_{ijl}$ in the $l$th layer (i.e., the matrix
$\mathbf{A}^{l}$). Other viable connectional null models include ones that
preserve degree \cite{Maslov2002,NG2004} or strength \cite{Ansmann2011}
distributions, or --- for networks based on time-series data --- preserve
length, frequency content, and amplitude distribution of the original time
series \cite{Bialonski2011}.  In this section, we present results for a few
null models that are applicable to a variety of temporal networks.  We note,
however, that this is a fruitful area of further investigation.

We employ two connectional null models specific for the broad classes of
networks represented by the brain and behavioral networks that we use as
examples in this paper. The brain networks provide an example of
time-dependent similarity networks, which are weighted and either fully
connected or almost fully connected \cite{Onnela2011}.  (The brain networks
have some $0$ entries in their corresponding adjacency tensors because we
have removed edges with weights that are not statistically significant
\cite{Bassett2011b}.)  We therefore employ a constrained null model that is
constructed by randomly rewiring edges while maintaining the empirical degree
distribution \cite{Maslov2002}. In Fig.~\ref{Fig4}A1, we demonstrate the use
of this null model to assess dynamic community structure. Importantly, this
constrained null model can be used in principle for any binary or weighted
network, though it does not take advantage of specific structure (aside from
strength distribution) that one might want to exploit. For example, the
behavioral networks have chain-like topologies, and it is desirable to
develop models that are specifically appropriate for such situations.  (One
can obviously make the same argument for other specific topologies.)  We
therefore introduce a highly constrained connectional null model that is
constructed by reassigning edge weights uniformly at random to existing
edges.  This does not change the underlying binary topology. (That is, we
preserve network topology but scramble network geometry.) We demonstrate the
use of this null model in Fig.~\ref{Fig4}B1.

In addition to intra-layer null models, one can also employ inter-layer null
models --- such as ones that scramble time or node identities
\cite{Bassett2011b}. For example, we construct  a \emph{temporal null model}
by randomly permuting the order of the network layers.  This temporal null
model can be used to probe the existence of significant temporal evolution of
community structure. One can also construct a \emph{nodal null model} by
randomly permuting the inter-layer edges that connect nodes in one layer to
nodes in another.  After the permutation is applied, an inter-layer edge can,
for example, connect node $i$ in layer $t$ with node $j \neq i$ in layer
$t+1$ rather than being constrained to connect each node $i$ in layer $t$
with itself in layer $t+1$. One can use this null model to probe the
importance of node identity in network organization.  We demonstrate the use
of our temporal null model in row 2 of Fig.~\ref{Fig4}, and we demonstrate
the use of our nodal null model in row 3 of Fig.~\ref{Fig4}.

%%%%%%%%%%%%

\subsubsection{Calculation of Diagnostics on Real Versus Null-Model Networks}

We characterize the effects of post-optimization null models using four
diagnostics: maximized modularity $Q$, the number of communities $n$, the
mean community size $s$, and the stationarity $\zeta$ (see the section titled `Network Diagnostics' for definitions). Due to the possibly large number of partitions
with nearly optimal $Q$ \cite{Good2010}, the values of such
diagnostics vary over realizations of a computational heuristic for both
the real and null-model networks.  (We call this \emph{optimization
variance}.)  The null-model networks also have a second source of variance
(which we call \emph{randomization variance}) from the myriad possible
network configurations that can be constructed from a randomization
procedure. We note that a third type of variance --- \textit{ensemble
variance} --- can also be present in systems containing multiple networks.  In the example data sets that we discuss, this represents variability among experimental subjects.

\begin{figure*}[]
\includegraphics[width=1\textwidth]{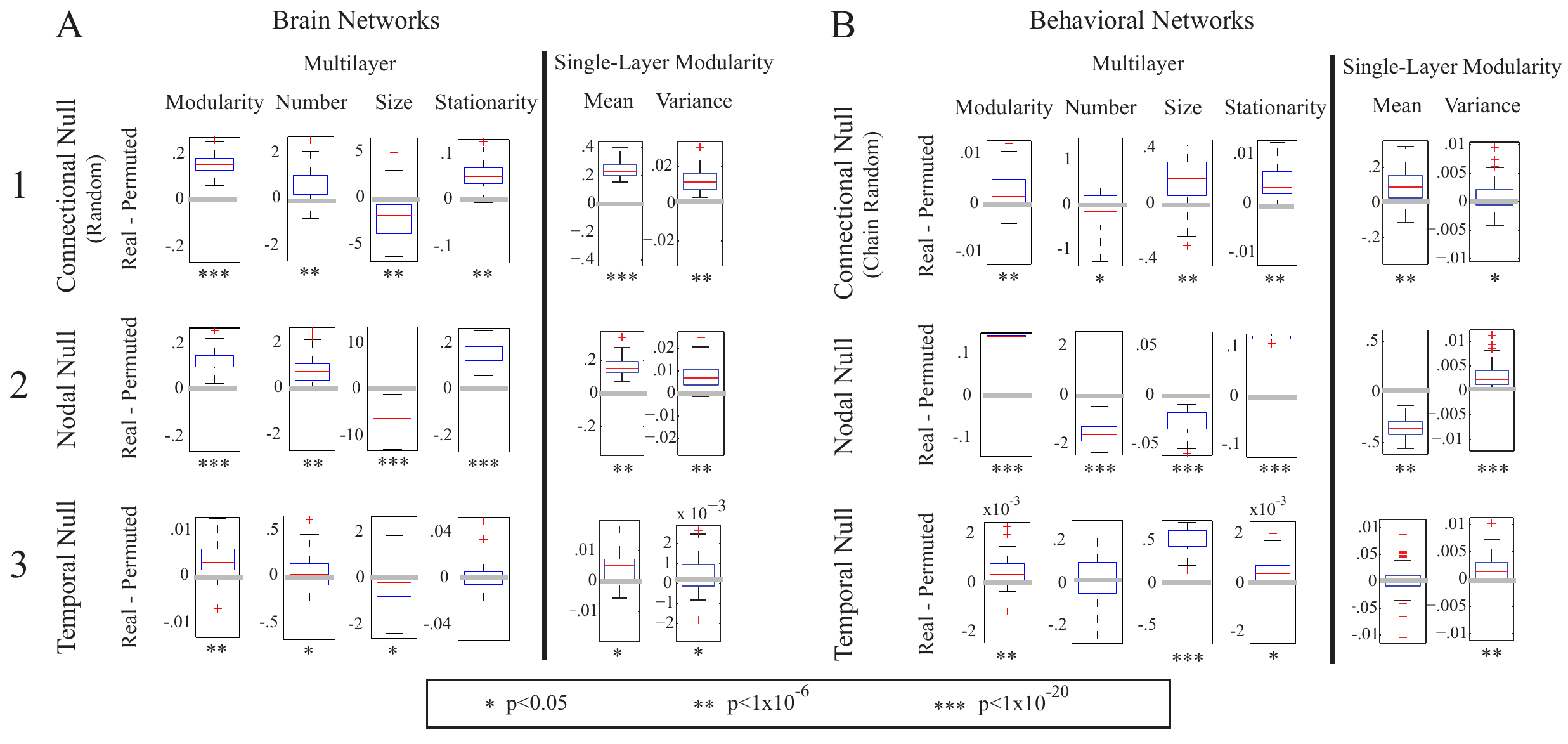}
\caption[]{(Color Online) Post-optimization null models. We compare four
multilayer diagnostics (optimized modularity, number of communities, mean
community size, and stationarity) and two single-layer diagnostics (mean and
variance of $Q_{s}$) for (\emph{A}) brain and (\emph{B}) behavioral networks
with the connectional (row 1), nodal (row 2), and temporal (row 3) null-model
networks. Box plots indicate quartiles and 95\% confidence intervals over the
individuals and experimental conditions. The structural resolution parameter
is $\gamma = 1$ and the temporal resolution parameter is $\omega = 1$.
\label{Fig4} }
\end{figure*}

We test for statistical differences between the real and null-model networks
as follows. We first compute $C=100$ optimizations of the modularity quality
function for a network constructed from real data and then compute the mean
of each of the four diagnostics over these $C$ samples. This yields
representative values of the diagnostics. We then maximize modularity for $C$
different randomizations of a given null model (i.e., 1 optimization per
randomization) and then compute the mean of each of the four diagnostics over
these $C$ samples. For both of our example data sets, we perform this
two-step procedure for each network in the ensemble (60 brain networks and 66
behavioral networks; see `Methods'). We then investigate whether the set of
representative diagnostics for the networks constructed from real data are
different from those of appropriate ensembles of null-model networks. To
address this issue, we subtract the diagnostic value for the null model from
that of the real network for each subject and experimental session. We then
use one-sample $t$-tests to determine whether the resulting distribution
differs significantly from $0$.  We show our results in Fig.~\ref{Fig4}.

Results depend on all three factors (the data set, the null model, and the
diagnostic), but there do seem to be some general patterns. For example, the
real networks exhibit the most consistent differences from the nodal null
model for all diagnostics and both data sets (see row 2 of Fig.~\ref{Fig4}).
For both data sets, the variance of single-layer modularity in the real
networks is consistently greater than those for all three null models,
irrespective of the mean (see the final two columns of Figs.~\ref{Fig4}A,B);
this is a potential indication of the statistical significance of the
temporal evolution. However, although optimized modularity is higher in the
real network for both data sets, the number of communities is higher in the
set of brain networks and lower in the set of behavioral networks. Similarly,
in comparison to the connectional null model, higher modularity is associated
with a smaller mean community size in the brain networks but a larger mean
size in the behavioral networks (see row 1 of Fig.~\ref{Fig4}). These results
demonstrate that the three post-optimization null models provide different
information about the network structure of the two systems and thereby
underscores the critical need for further investigations of null-model
construction.

%%%%%%%%%%

\subsection{Structural and Temporal Resolution Parameters}

When optimizing multilayer modularity, we must choose (or otherwise derive) values for the
structural resolution parameter $\gamma$ and the temporal resolution
parameters $\omega$. By varying $\gamma$, one can tune the size of
communities within a given layer: large values of $\gamma$ yield more
communities, and small values yield fewer communities.  A systematic method
for how to determine values of $\omega_{jlr}$ has not yet been discussed in the
literature. In principle, one could choose different $\omega_{jlr}$ values
for different nodes, but we focus on the simplest scenario in which the value
of $\omega_{jlr} \equiv \omega$ is identical for all nodes $j$ and all
contiguous pairs of layers $l$ and $r$ (and is otherwise $0$). In this framework, the temporal
resolution parameter $\omega$ provides a means of tuning the number of
communities discovered across layers: high values of $\omega$ yield fewer
communities, and low values yield more communities. It is beneficial to study
a range of parameter values to examine the breadth of structural (i.e.,
intra-layer \cite{Fortunato2007,Good2010,Traag2011}) and temporal (i.e., inter-layer)
resolutions of community structure, and some papers have begun to make
progress in this direction
\cite{Onnela2011,Mucha2010,Bassett2011b,Macon2012,Wymbs2011}.

To characterize community structure as a function of resolution-parameter
values (and hence of system scales), we quantify the quality of partitions
using the mean value of optimized $Q$.  To do this, we examine the constitution of the partitions
using the mean similarity over $C$ optimizations, and we compute partition
similarities using the $z$-score of the Rand coefficient
\cite{Traud2010}. For comparing two partitions $\alpha$ and $\beta$,
we calculate the Rand $z$-score in terms of the network's total number of pairs of
nodes $M$, the number of pairs $M_{\alpha}$ that are in the
same community in partition $\alpha$, the number of pairs $M_{\beta}$ that
are in the same community in partition $\beta$, and the number of pairs $w_{\alpha \beta}$ that are assigned to the same community both in
partition $\alpha$ and in partition $\beta$. The $z$-score of the Rand
coefficient comparing these two partitions is
\begin{equation}
	z_{\alpha\beta} = \frac{1}{\sigma_{w_{\alpha \beta}}}
	\left(w_{\alpha \beta}-\frac{M_{\alpha}M_{\beta}}{M}\right)\,,
\end{equation}
where $\sigma_{w_{\alpha \beta}}$ is the standard deviation of $w_{\alpha
\beta}$ (as in \cite{Traud2010}). Let the \emph{mean partition similarity} $z$ denote the mean value
of $z_{\alpha \beta}$ over all possible partition pairs for $\alpha \neq
\beta$.

\begin{figure*}[]
\includegraphics[width=.8\textwidth]{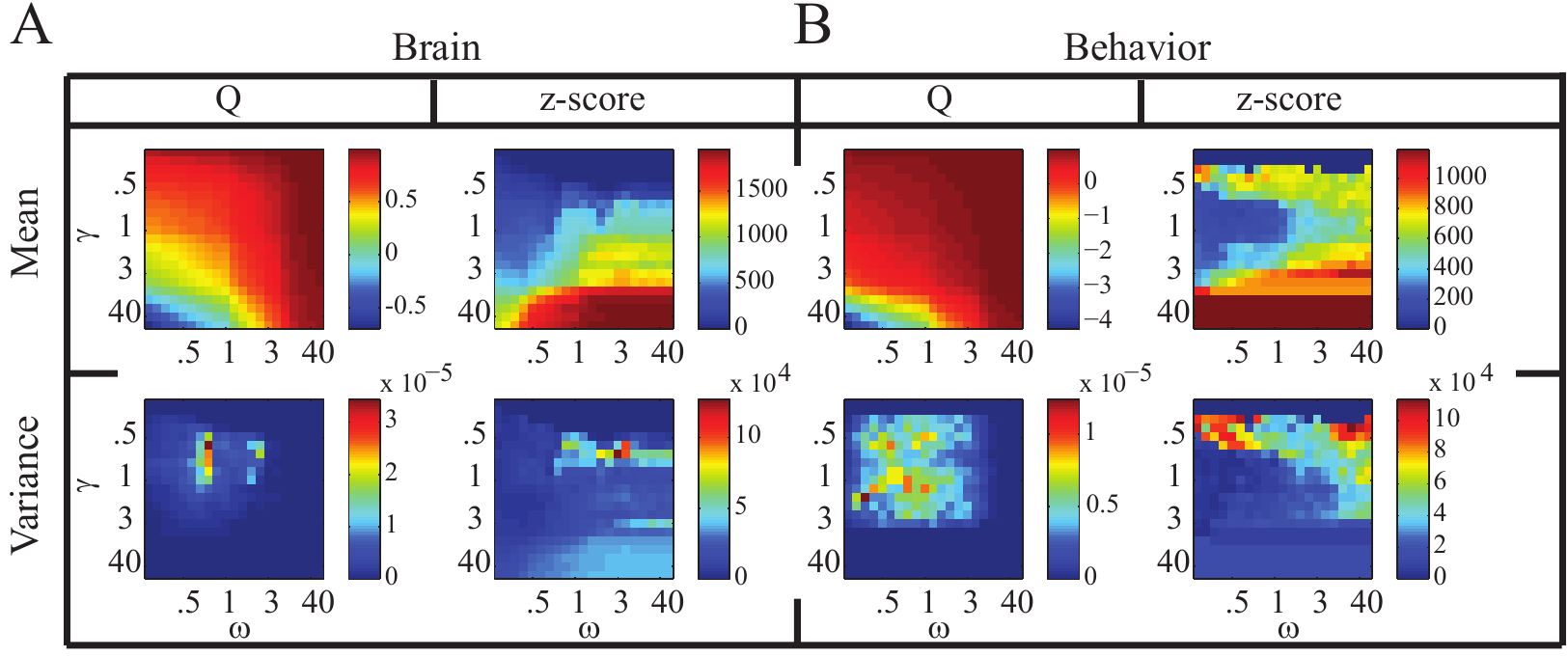}
\caption[]{(Color Online) Optimized modularity $Q$ and Rand $z$-score as
functions of the resolution parameters $\gamma$ and $\omega$ for the
(\emph{A}) brain and (\emph{B}) behavioral networks. The top row shows the
mean value of maximized $Q$ over $C = 100$ optimizations and the mean
partition similarity $z$ over all possible pairs of the $C$ partitions. The
bottom row shows the variance of maximized $Q$ over the optimizations and the
variance of the partition similarity over all possible pairs of partitions.
The results shown in this figure come from a single individual and
experimental scan, but we obtain qualitatively similar results for other
individuals and scans. Note that the axis scalings are nonlinear.
\label{Fig5}}
\end{figure*}

In Fig.~\ref{Fig5}, we show both $z$ and optimized $Q$ as a function of
$\gamma$ and $\omega$ in both brain and behavioral networks. The highest
modularity values occur for low $\gamma$ and high $\omega$. The mean
partition similarity is high for large $\gamma$ in the brain networks, and it
is high for both small and large $\gamma$ in the behavioral networks.
Interestingly, in both systems, the partition similarity when
$\gamma=\omega=1$ is lower than it is elsewhere in the $(\gamma,\omega)$
parameter plane, so the variability in partitions tends to be large at this
point. Indeed, as shown in the second row of Fig.~\ref{Fig5}, modularity
exhibits significant variability for $\gamma=\omega=1$ compared to other
resolution-parameter values.

It is useful to be able to determine the ranges of $\gamma$
and $\omega$ that produce community structure that is significantly different
from a particular null model. One can thereby use null models to probe
resolution-parameter values at which a network displays interesting structures. This
could be especially useful in systems for which one wishes to identify length
scales (such as a characteristic mean community size) or time scales
\cite{danigranular,Bassett2011b,Palla2007,Lou2011} directly from data.

\begin{figure*}[]
\includegraphics[width=.8\textwidth]{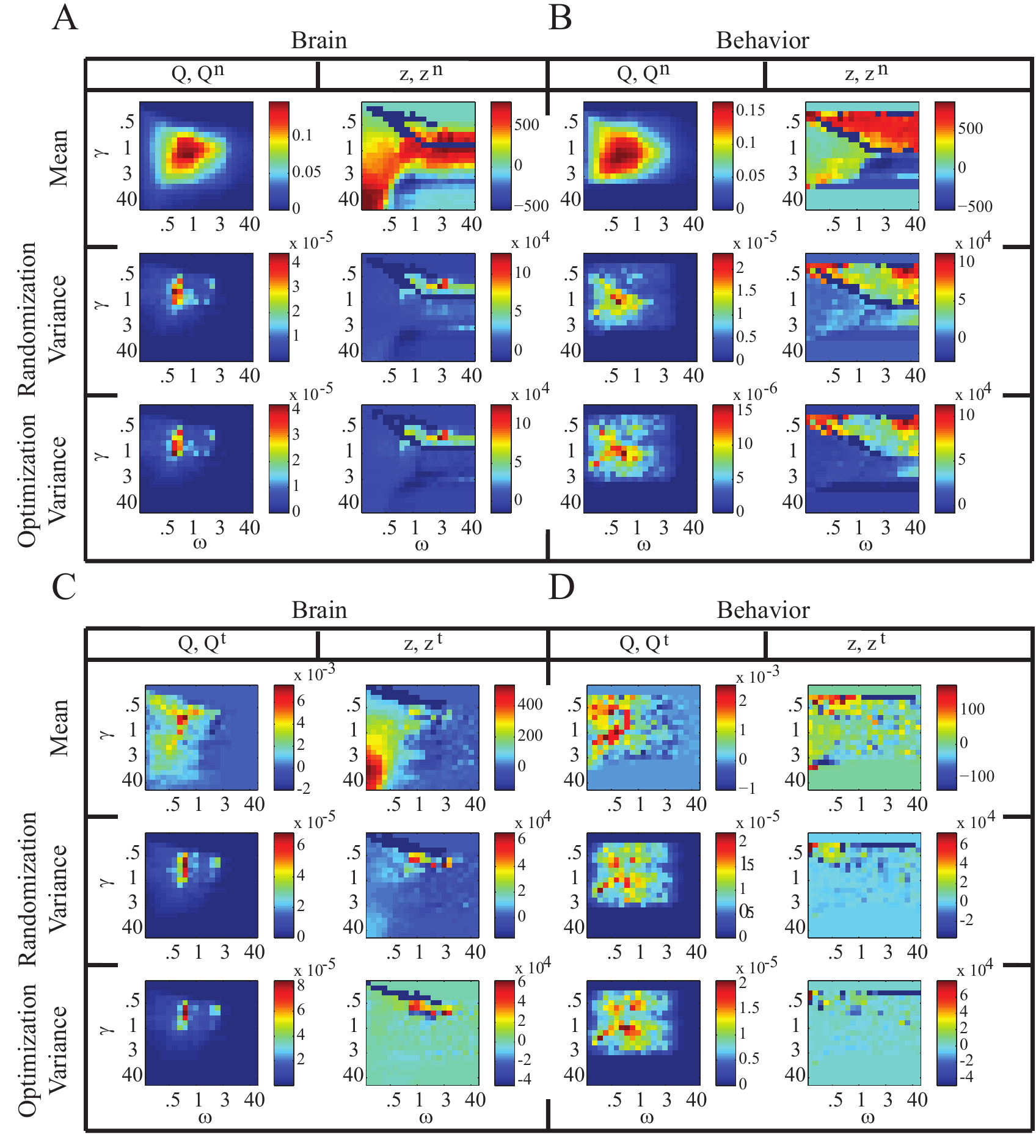}
\caption[]{(Color Online) Differences, as a function of $\gamma$ and
$\omega$, between the real networks and the (\emph{A,B}) nodal and
(\emph{C,D}) temporal null models for maximized modularity $Q$ and partition
similarity $z$ for the (\emph{A,C}) brain and (\emph{B,D}) behavioral
networks. The first row in each panel gives the difference in the mean values
of the diagnostic variables between the real and null-model networks.  Panels
(\emph{A,B}) show the results for $Q-Q^{n}$ and $z-z^{n}$, and panels
(\emph{C,D}) show the results for $Q-Q^{t}$ and $z-z^{t}$. The quantities $Q$
and $z$ again denote the modularity and partition similarity of the real
network, $Q^{n}$ and $z^{n}$ denote the modularity and partition similarity
of the nodal null-model network, and $Q^{t}$ and $z^{t}$ denote the
modularity and partition similarity of the temporal null-model network. The
second row in each panel gives the difference between the optimization
variance of the real network and the randomization variance of the null-model
network for the same diagnostic variable pairs. The third row in each panel
gives the difference in the optimization variance of the real network and the
optimization variance of the null-model network for the same diagnostic
variable pairs. We show results for a single individual and scan in the
experiment, but results are qualitatively similar for other individuals and
scans. Note that the axis scalings are nonlinear. \label{Fig6}}
\end{figure*}

In Fig.~\ref{Fig6}, we show examples of how the difference between diagnostic
values for real and null-model networks varies as a function of $\gamma$ and
$\omega$.  As illustrated in panels (A) and (B), the brain and behavioral
networks both exhibit a distinctly higher mean optimized modularity than the
associated nodal null-model network for $\gamma \approx \omega \approx 1$.
Interestingly, this roundly peaked difference in $Q$ is not evident in
comparisons of the real networks to temporal null-model networks (see
Figs.~\ref{Fig6}C,D), so resolution-parameter values (and hence system scales)
of potential interest might be more identifiable by comparison to nodal than
to temporal null models in these examples. It is possible, however, that
defining temporal layers over a longer or shorter duration would yield
identifiable peaks in the difference in $Q$.

The differences in the Rand $z$-score landscapes are more difficult to
interpret, as the values of mean partition similarity $z$ are much larger in
the real networks for some resolution-parameter values (positive differences;
red) but are much larger in the null-model networks for other
resolution-parameter values (negative differences; blue).   The clearest
situation occurs when comparing the brain's real and temporal null-model
networks (see Fig.~\ref{Fig6}C), as the network built from real data exhibits
a much larger value of $z$ (and hence much more consistent optimization
solutions) than the temporal null-model networks for high values of $\gamma$
(i.e., when there many communities) and low $\omega$ (i.e., when there is
weak temporal coupling). These results are consistent with the fact that weak
temporal coupling in a multilayer network facilitates greater temporal
variability in network partitions across time. Such variability appears to be
significantly different than the noise induced by scrambling time layers.
These results suggest potential resolution values of interest for the brain
system, as partitions are very consistent across many optimizations. For
example, it would be interesting to investigate community structure in these
networks for high $\gamma$ (e.g., $\gamma \approx 40$) and low $\omega$
(e.g., $\omega \approx$ 0.1). At these resolution values, one can identify
smaller communities with greater temporal variability than the communities
identified for the case of $\gamma=\omega=1$ \cite{Bassett2011b}.

The optimization and randomization variances appear to be similar in
the brain and behavioral networks (see rows 2--3 in every panel of
Fig.~\ref{Fig6}) not only in terms of their mean values but also in their distribution in
the part of the $(\gamma,\omega)$ parameter plane that we examined. In particular, the variance
in $Q$ is larger in the real networks precisely where the mean is also
larger, so mean and variance are likely either dependent on one another or on
some common source.  Importantly, such dependence influences the ability to draw statistical
conclusions because it is possible that the points in the
$(\gamma,\omega)$ plane with the \emph{largest} differences in mean are not
necessarily the points with the \emph{most significant} differences in mean.

We also find that the dependencies of the diagnostics on $\gamma$ and $\omega$ are
consistent across subjects and scans, suggesting that our results are
ensemble-specific rather than individual-specific.

%%%%%%%%%%%%%

\subsection{Examination of Data Generated from a Dynamical System}
Real-world data is often clouded by unknown or mathematically undefinable
sources of variance, so it is also important to examine data sets generated
from dynamical systems (or other models). Because we are concerned with
time-dependent networks, we consider an example consisting of time-dependent
data generated by a well-known dynamical system.

We construct a network of Kuramoto oscillators, in which the phase
$\theta_{i}(t)$ of the $i^{\mathrm{th}}$ oscillator evolves in time according
to
\begin{equation}
	\frac{d \theta_{i}}{dt} = \omega_{i} + \sum_{j} \kappa A_{ij} \mathrm{sin}(\theta_{j}-\theta_{i})\,, \quad i \in \{1, \ldots, N\}\,,
\end{equation}
where $\omega_{i}$ is the natural frequency of oscillator $i$, the matrix
$\mathbf{A}$ gives the binary coupling between each pair of oscillators, and
$\kappa$ is a positive real constant that indicates the strength of the
coupling. We draw the frequencies $\omega_i$ from a Gaussian distribution
with mean $0$ and standard deviation $1$. In our simulations, we use a time
step of $\tau=0.1$, a constant of $\kappa=0.2$, and a network size of
$N=128$.

\begin{figure}[]
\includegraphics[width=.45\textwidth]{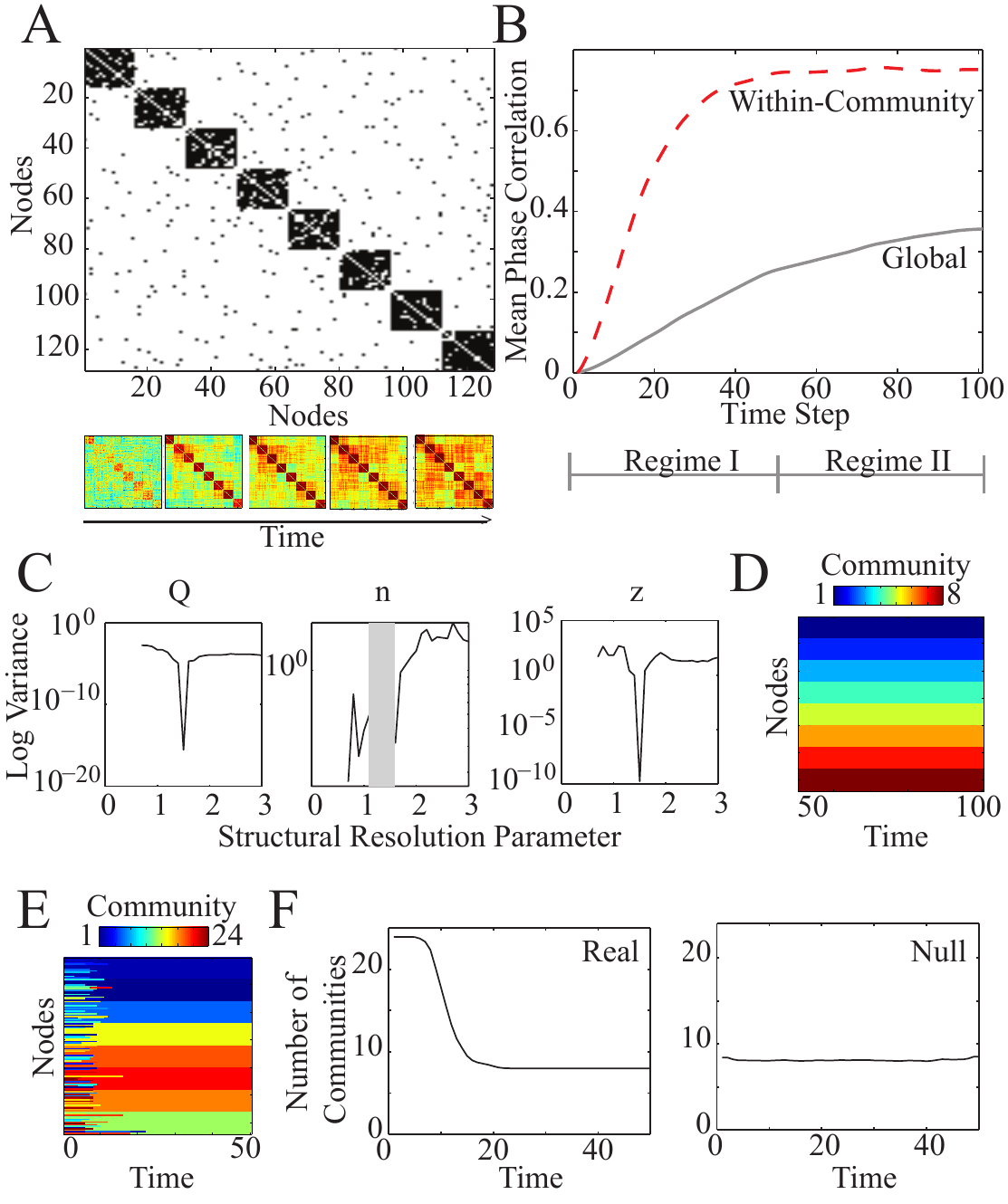}
\caption[]{(Color Online) Dynamic community detection in a
network of Kuramoto oscillators. (\emph{A, top}) The coupling matrix between
$N=128$ phase oscillators contains 8 communities, each of which has 16
nodes. (\emph{A, bottom}) Over time, oscillators synchronize with one
another. Color indicates the mean phase correlation between oscillators,
where hotter (darker gray) colors indicate stronger
correlations. (\emph{B}) Phase correlation between oscillators as a function
of time. The mean phase correlation between oscillators in the same community
(dashed red curve) increases faster than the mean phase correlation between
all oscillators in the system (solid gray curve). Regime I encompasses the
first 50 time steps, and regime II emcompasses the subsequent 50 time steps.
(\emph{C}) Variance of maximized multilayer modularity (left), number of
communities (middle), and partition similarity $z$ (right) over $100$
optimizations of the multilayer modularity quality function for the temporal
network in regime II as a function of the structural resolution parameter for
$\omega=1$. The shaded gray area indicates values of the structural
resolution parameter that provide $0$ variance in the number of communities.
(\emph{D}) Example partition of the temporal network in regime II at
$\gamma=1.5$, which occurs near the troughs in panel (\emph{C}). (\emph{E})
Example partition of the temporal network in regime I at $\gamma=1.5$.
(\emph{F}) Number of communities as a function of time for (\emph{left}) the
temporal network in regime I and (\emph{right}) its corresponding temporal
null model. We averaged values over $C = 100$ optimizations of multilayer
modularity.
 \label{fig:kuramoto}}
\end{figure}

Kuramoto oscillators have been studied in the context of various network
topologies and geometries
\cite{Arenasreview08,Kuramoto2002,Shima2004,Abrams2004,Arenas2006,Stout2011}
and from both the component and ensemble perspectives \cite{Barlev2011}. We
are interested in networks with dynamic community structure. Following
Refs.~\cite{Arenas2006,Arenas2006b}, we impose a well-defined community
structure in which each community is composed of $16$ nodes. In each time
step, each node has $13$ connections with nodes in its own community and $1$
connection with nodes outside of its community (see
Fig.~\ref{fig:kuramoto}A).

To quantify the temporal evolution of synchronization patterns, we define a
set of temporal networks given by the time-dependent correlation between
pairs of oscillators:
\begin{equation}
	\phi_{ij}(t) = \left\langle |\cos[\theta_{i}(t) - \theta_{j}(t)]| \right\rangle\,,
\end{equation}
where the angular brackets indicate an average over $20$ simulations. As time
evolves from time step $t=0$ to $t=100$, oscillators tend to synchronize with
other oscillators in their same community more quickly than with oscillators
in other communities (see Fig.~\ref{fig:kuramoto}B).

To examine the performance of our multilayer community-detection techniques
in this example, we compute $A_{ijl} = A_{ijt} = \phi_{i,j}(t)$ and using the
multilayer extension of the Newman-Girvan null model $P_{ijl}$ given in
Eq.(~\ref{pijl}). We separately optimize $Q$ for two temporal regimes: (1)
regime I (with $t \in \{1, \ldots, 50\}$), for which synchronization within
communities increases rapidly; and (2) regime II (with $t \in \{51, \ldots,
100\}$), for which within-community synchronization level is roughly constant
but global synchronization still increases gradually. We set $\omega=1$ and
probe the effects of the structural resolution parameter $\gamma$ in regime
II. In Figs.~\ref{fig:kuramoto}C,D, we illustrate that one can identify the
value of $\gamma$ that best uncovers the underlying hard-wired connectivity
using troughs in the optimization variance of several diagnostics (e.g.,
maximized modularity, number of communities, and mean partition similarity).

We probe the community structure in regime I using the value of $\gamma$ that
best uncovered the underlying hard-wired connectivity in regime II. We
observe temporal changes of community structure at early time points, as
evidenced by the large number of communities for $t \in \{1, \ldots, 5\}$
(see Figs.~\ref{fig:kuramoto}E,F). Importantly, the temporal dependence of
community number on $t$ is not expected from a post-optimization temporal
null model (see the right panel of Fig.~\ref{fig:kuramoto}F). We obtain
qualitatively similar results when we optimize the multilayer modularity
quality function over the entire temporal network without separating the data
into two regimes.

Our results illustrate that one can use dynamic community detection to
uncover the resolution of inherent hard-wired structure in a data set
extracted from the temporal evolution of a dynamical system and that
post-optimization null models can be used to identify regimes of unexpected
temporal dependence in network structure.

%%%%%%%%%%%%%%

\subsection{Dealing With Degeneracy: Constructing Representative Partitions}

\begin{figure}[]
\includegraphics[width=.40\textwidth]{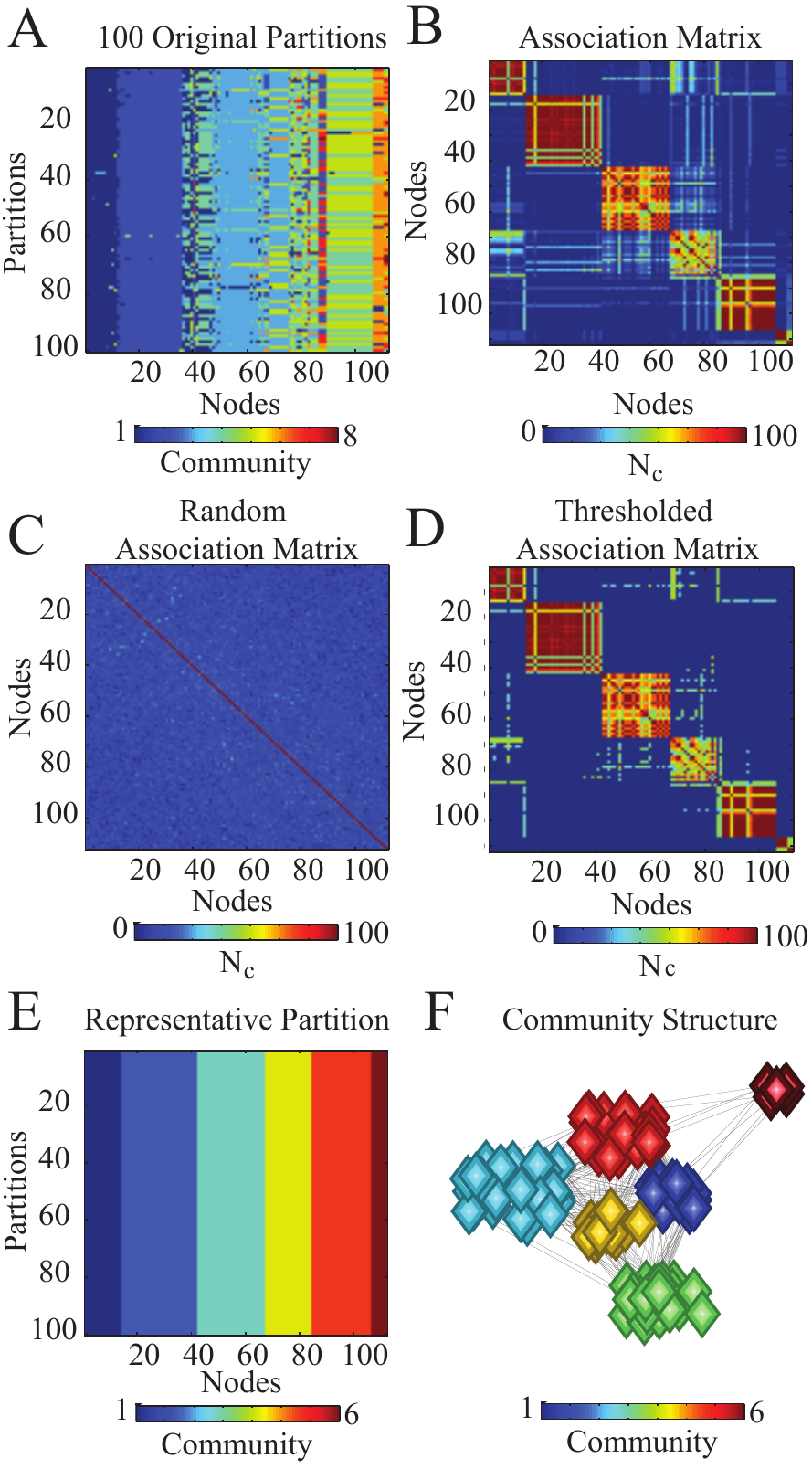}
\caption[]{(Color Online) Constructing representative partitions for an
example brain network layer. (\emph{A}) Partitions extracted during $C$
optimizations of the quality function $Q$. (\emph{B}) The $N \times N$ nodal
association matrix $\mathbf{T}$, whose elements indicate the number of times
node $i$ and node $j$ have been assigned to the same community. (\emph{C})
The $N \times N$ random nodal association matrix $\mathbf{T^{r}}$, whose
elements indicate the number of times node $i$ and node $j$ are expected to
be assigned to the same community by chance. (\emph{D}) The thresholded nodal
association matrix $\mathbf{T'}$, where elements with values less than those
expected by chance have been set to $0$. (\emph{E}) Partitions extracted
during $C = 100$ optimizations of the single-layer modularity quality
function $Q_{s}$ for the matrix $\mathbf{T}$ from panel (\emph{D}). Note that
each of the $C$ optimizations yields the same partition. (\emph{F})
Visualization of the representative partition given in (\emph{E})
\cite{Traud2009}. We have reordered the nodes in the matrices in panels
(\emph{A}-\emph{E}) for better visualization of community structure.
\label{Fig7}}
\end{figure}

The multilayer modularity quality function has numerous near-degeneracies, so
it is important to perform many instantiations when using a non-deterministic
computational heuristic to optimize modularity \cite{Good2010}. In doing
this, an important issue is how (and whether) to distill a single
representative partition from a (possibly very large) set of $C$ partitions
\cite{Lancichinetti2012}. In Fig.~\ref{Fig7}, we illustrate a new method for
constructing a representative partition based on statistical testing in
comparison to null models.

Consider $C$ partitions from a single layer of an example multilayer brain
network (see Fig.~\ref{Fig7}A). We construct a nodal association matrix
$\mathbf{T}$, where the element $T_{ij}$ indicates how many times nodes $i$
and $j$ have been assigned to the same community (see Fig.~\ref{Fig7}B). We
then construct a null-model association matrix $\mathbf{T^{r}}$ based on
random permutations of the original partitions (see Fig.~\ref{Fig7}C). That
is, for each of the $C$ partitions, we reassign nodes uniformly at random to
the $n$ communities of mean size $s$ that are present in the selected
partition. For every pair of nodes $i$ and $j$, we let $T_{ij}^{r}$ be the
number of times these two nodes have been assigned to the same community in
this permuted situation (see Fig.~\ref{Fig7}C). The values $T_{ij}^{r}$ then
form a distribution for the expected number of times two nodes are assigned
to the same partition. Using an example with $C=100$, we observe that two
nodes can be assigned to the same community up to about $30$ times out of the
$C$ partitions purely by chance. To be conservative, we remove such `noise'
from the original nodal association matrix $\mathbf{T}$ by setting any
element $T_{ij}$ whose value is less than the maximum entry of the random
association matrix to $0$ (see Fig.~\ref{Fig7}D).  This yields the
thresholded matrix $\mathbf{T'}$, which retains statistically significant
relationships between nodes.

We use a Louvain-like algorithm to perform $C$ optimizations of the
single-layer modularity $Q_0$ for the thresholded matrix $\mathbf{T'}$.
Interestingly, this procedure typically extracts identical partitions for
each of these optimizations in our examples (see Fig.~\ref{Fig7}E). This
method therefore allows one to deal with the inherent near-degeneracy of the
modularity quality function and provides a robust, representative partition
of the original example brain network layer (see Fig.~\ref{Fig7}F).

\begin{figure}[]
\includegraphics[width=.4\textwidth]{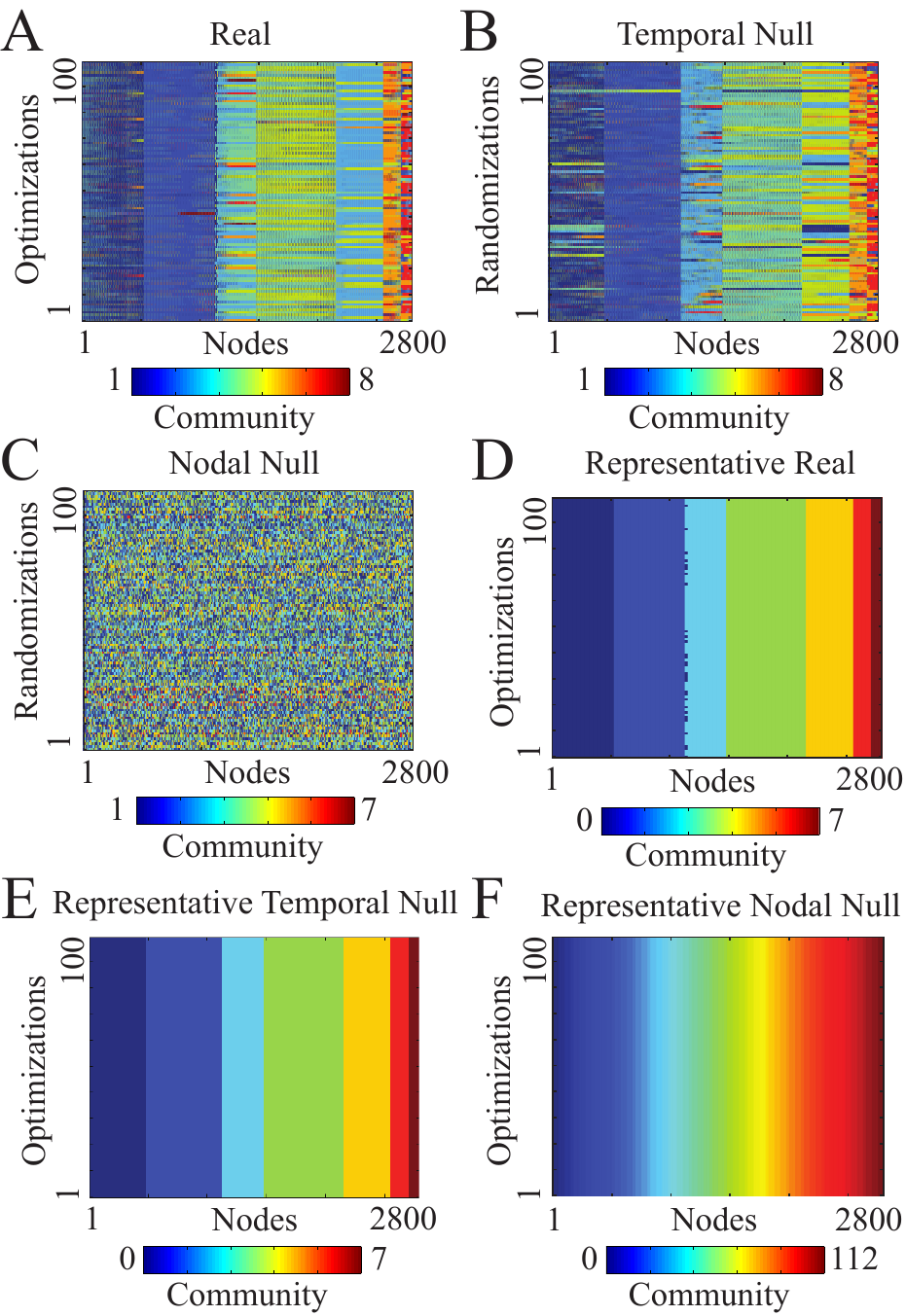}
\caption[]{(Color Online) Representative partitions of multilayer brain
networks for an example subject and scan. (\emph{A}) Partitions extracted
for $C = 100$ optimizations of the quality function $Q$ on the real
multilayer network (112 nodes $\times$ 25 time windows, which yields 2800
nodes in total). Partitions extracted for $C$ randomizations for the
(\emph{B}) temporal and (\emph{C}) nodal null-model networks.  (\emph{D})
Partitions extracted for $C$ optimizations of the quality function $Q$ of
the thresholded nodal association matrix for the (\emph{D}) real, (\emph{E})
temporal null-model, and (\emph{F}) nodel null-model networks. Note that the
partitioning is robust to multiple optimizations.  We have reordered the
nodes in each column for better visualization of community structure.
The structural resolution parameter is $\gamma=1$, and the temporal resolution
parameter is $\omega=1$.
\label{Fig8}}
\end{figure}

We apply the same method to multilayer networks (see Fig.~\ref{Fig8}) to find
a representative partition of (1) a real network over $C$ optimizations, (2)
a temporal null-model network over $C$ randomizations, and (3) a nodal
null-model network over $C$ randomizations. Using these examples, we have
successfully uncovered representative partitions when they appear to exist
(e.g., in the real networks and the temporal null-model networks) and have
not been able to uncover a representative partition when one does not appear
to exist (e.g., in the nodal null-model network, for which each of the 112
brain nodes is placed in its own community in the representative partition).
We also note that the representative partitions in the temporal null-model
and real networks largely match the original data in terms of both sizes and
number of communities. These results indicate the potential of this method to
uncover meaningful representative partitions over optimizations or
randomizations in multilayer networks.

%%%%%%%%%%%%%

\section{Conclusions}

In this paper, we discussed methodological issues in the determination and
interpretation of dynamic community structure in multilayer networks.  We
also analyzed the behavior of several null models used for optimizing quality
functions (such as modularity) in multilayer networks.

We described the construction of networks and the effects that certain
choices can have on the use of both optimization and post-optimization null
models. We introduce novel modularity-optimization null models for two cases:
(1) networks composed of ordered nodes (a `chain null model') and (2)
networks constructed from time-series similarities (FT and AAFT surrogates).
We studied `connectional', `temporal', and `nodal' post-optimization null
models using several multilayer diagnostics (optimized modularity, number of
communities, mean community size, and stationarity) as well as novel
single-layer diagnostics (in the form of measures based on time series for
optimized modularity).  To investigate the utility of such considerations for
model-generated data, we also applied our methodology to time-series data
generated from coupled Kuramoto oscillators.

We examined the dependence of optimized modularity and partition similarity on
structural and temporal resolution parameters as well as the influence of their
variances on putative statistical conclusions.  Finally, we described a
simple method to address the issue of near-degeneracy in the landscapes of
modularity and similar quality functions using a method to construct a
robust, representative partition from a network layer.

The present paper illustrates what we feel are important considerations in
the detection of dynamic communities.  As one considers data with increasingly
complicated structures, network analyses of such data must become
increasingly nuanced, and the purpose of this paper has been to discuss and
provide some potential starting points for how to try to address some of
these nuances.

%%%%%%%%%

\section*{Acknowledgments}

We thank L.~C. Bassett, M. Bazzi, K. Doron, L. Jeub, S.~H. Lee, D. Malinow,
and the anonymous referees for useful discussions. This work was supported by
the Errett Fisher Foundation, the Templeton Foundation, David and Lucile
Packard Foundation, PHS Grant NS44393, Sage Center for the Study of the Mind,
and the Institute for Collaborative Biotechnologies through contract no.
W911NF-09-D-0001 from the U.S. Army Research Office.  MAP was supported by
the James S. McDonnell Foundation (research award \#220020177), the EPSRC
(EP/J001759/1), and the FET-Proactive project PLEXMATH (FP7-ICT-2011-8; grant
\#317614) funded by the European Commission.  He also acknowledges SAMSI and
KITP for supporting his visits to them. PJM acknowledges support from Award
Number R21GM099493 from the National Institute of General Medical Sciences;
the content is solely the responsibility of the authors and does not
necessarily represent the official views of the National Institute of General
Medical Sciences or the National Institutes of Health.

%%%%%%%

%\bibliographystyle{apsrev4-1_CompAuList} % style file which compresses author lists for >4 authors
\bibliographystyle{apsrev4-1} % Don't compress author lists
\bibliography{bibfile7}% Produces the bibliography via BibTeX.

%\noindent {\bf Fig.~1.} Fig.~1 caption for submission

\end{document}